\documentclass[authoryear,preprint,review,12pt]{elsarticle}

\usepackage{amssymb}
\usepackage{amsthm}

\usepackage{lineno}

\usepackage[utf8]{inputenc}
\usepackage{csvsimple}
\usepackage{gensymb} 
\usepackage{amsmath}
\usepackage{float}
\usepackage{natbib}
\usepackage{color, colortbl}
\usepackage{geometry}
\usepackage[justification=centering]{caption}
\usepackage{subcaption} 

\definecolor{Gray}{gray}{0.9}
\newcommand{\Hugzero}{\mbox{\normalfont\Huge\bfseries 0}}
\newcommand{\bigsigma}{\mbox{\normalfont\Huge\bfseries $\Sigma$}}

\journal{Icarus}

\begin{document}

\begin{frontmatter}{}

\title{Regional study of Europa’s photometry}

\author[label1]{I. Belgacem\corref{cor1}}
\ead{ines.belgacem@u-psud.fr}

\author[label1]{F. Schmidt}
\author[label2]{G. Jonniaux}

\address[label1]{GEOPS, Univ. Paris-Sud, CNRS, Université Paris-Saclay, 91405 Orsay, France. }
\address[label2]{Airbus Defence $\&$ Space, Toulouse, France.}

\cortext[cor1]{Corresponding author}

\begin{abstract}
The surface of Europa is geologically young and shows signs of current activity. Studying it from a photometric point of view gives us insight on its physical state. We used a collection of 57 images from Voyager's Imaging Science System and New Horizons' LOng Range Reconnaissance Imager for which we corrected the geometric metadata and projected every pixel to compute photometric information (reflectance and geometry of observation). We studied 20 areas scattered across the surface of Europa and estimated their photometric behavior using the Hapke radiative transfer model and a Bayesian framework in order to estimate their microphysical state. We have found that most of them were consistent with the bright backscattering behavior of Europa, already observed at a global scale, indicating the presence of grains maturated by space weathering. However, we have identified very bright areas showing a narrow forward scattering possibly indicating the presence of fresh deposits that could be attributed to recent cryovolcanism or jets. Overall, we showed that the photometry of Europa's surface is more diverse than previously thought and so is its microphysical state.
\end{abstract}


\end{frontmatter}{}


\section{Introduction}
Europa is a prime candidate in the search for habitability in our Solar System. The surface of the moon is the youngest of the Jovian icy satellites and shows signs of recent and current geological activity \citep{Kattenhorn_EvidenceSubductionIce_NG_2014}. This activity may be driven by the presence of a global ocean underneath the icy surface \citep{Pappalardo_DoesEuropaHave_JoGRP_1999} made of salt-rich water \citep{Zimmer_SubsurfaceOceansEuropa_I_2000, Kivelson_GalileoMagnetometerMeasurements_S_2000}. Recent studies showed exciting results finding possible plume eruptions \citep{Roth_TransientWaterVapor_S_2014,  Rhoden_LinkingEuropasPlume_I_2015,  Sparks_ProbingEvidencePlumes_AJ_2016} as well as the possibility for penitentes to form by sublimation of ice close to the equator \citep{Hobley_FormationMetreScale_NG_2018}. In this study, we use the photometry of Europa's surface to estimate regional physical properties to learn more about the surface processes at play in the icy satellite.

Significant work has been done in the past to study the integrated photometry of Europa using the Voyager spacecrafts images and telescopic observations \citep{Buratti_VoyagerPhotometryEuropa_I_1983, Buratti_ApplicationRadiativeTransfer_I_1985, Buratti_PhotometrySurfaceStructure_JoGR_1995, Domingue_EuropasPhaseCurve_I_1991,  Domingue_ReAnalysisSolar_I_1997}. \citealt{Domingue_DiskResolvedPhotometric_I_1992} looked at one particular region in the trailing hemisphere to study different types of terrains.  We want to expand that by proposing a regional study of 20 areas scattered across Europa's surface.

This study is also conducted in the context of space exploration by NASA's Europa Clipper and ESA's JUICE (JUpiter ICy moons Explorer) missions. The latter is scheduled to launch in 2022 and arrive at the Jovian system in 2030 to study Jupiter and its icy moons for three and a half years \citep{Grasset_JUpiterICymoons_PaSS_2013}. The spacecraft is being designed by Airbus Defence $\&$ Space in Toulouse, France, with a new and innovative navigation system \citep{Jonniaux_AutonomousVisionBased_IAC_2016}. To offer the best of that algorithm, the spacecraft needs to have a proper knowledge of the photometric models of the moons that will be observed.

Two pieces of information are needed to study the photometry: the geometry of observation and the reflectance. We have developed our own pipeline to process images taken by the Voyager spacecrafts and more recently by New Horizons to correct metadata errors such as spacecraft pointing. We have projected every individual pixel to derive the geometry of observation. The reflectance has been obtained by radiometric calibration of the images.

We are using a new approach based on a Bayesian framework to constrain the Hapke model parameters \citep{Hapke_TheoryReflectanceEmittance_book_1993} on selected areas of Europa. A similar approach has been used on CRISM data on Mars \citep{Fernando_SurfaceReflectanceMars_JoGRP_2013, Schmidt_RealisticUncertaintiesHapke_I_2015,  Fernando_MartianSurfaceMicrotexture_PaSS_2016, Schmidt_EfficiencyBRDFSampling_I_2019} and showed that a Bayesian inversion could give satisfying results on photometric data. 

By using the Hapke model, we will be looking at parameters that will give us insight on the physical properties of the surface in light of numerous experimental studies \citep{McGuire_ExperimentalStudyLight_I_1995, Domingue_ScatteringPropertiesNatural_I_1997, Cord_PlanetaryRegolithSurface_I_2003, Souchon_ExperimentalStudyHapkes_I_2011, Pommerol_PhotometryBulkPhysical_PaSS_2011, Jost_MicrometerSizedIce_I_2013, Jost_ExperimentalCharacterizationOpposition_I_2016} in the purpose of characterizing different regions of interest (ROIs) and highlighting compelling areas for future and more thorough observations.


\section{Dataset}

We use images from the Voyager probes dataset taken with the Imaging Science System (ISS) as well as from the New Horizons spacecraft taken with the LOng Range Reconnaissance Imager (LORRI). The complete list of the 57 images used in this study can be found in the supplementary material in table S1.

\subsubsection*{New Horizons' LORRI}
We retrieve the data from NASA's Planetary Data System (PDS) \citep{Cheng_PDS_NH_LORRI}. 

The images are partly calibrated and available in DN (Digital Number) units. To complete the calibration process, we use the process described in \citealt{Morgan_CalibrationNewHorizons_2005}. We compute the calibration coefficient for Europa using the spectra in \citealt{Calvin_SpectraIcyGalilean_JoGR_1995}. We obtain the value  $R_{Europa} = 2.7.10^{11} \frac{DN/s/pixel}{W/cm2/sr/nm}$. The radiance at the pivot wavelength of the instrument $\lambda_p = 607.6$ nm is then derived from the following equation \citep{Morgan_CalibrationNewHorizons_2005, Cheng_LongRangeReconnaissance_SSR_2008}:

\begin{equation}
\label{eq:calib_LORRI}
I = \frac{\frac{C}{T_{exp}}}{R_{Europa}}
\end{equation}

Where: 
\begin{itemize}
\item $I$: radiance at the pivot wavelength, in $W.s^{-1}.cm^{-2}.sr^{-1}.nm^{-1}$
\item $C$: pre-calibrated signal, in $DN/pix$
\item $T_{int}$: integration time, in $s$
\item $R_{Europa}$: spectral response of Europa in $\frac{DN.s^{-1}}{W.s^{-1}.cm^{-2}.sr^{-1}.nm^{-1}}$
\end{itemize}

To finally compute the reflectance, we use the solar spectrum \citep{Wehrli_ExtraterrestrialSolarSpectrum_PO+WRCDDS_1985} and compute the reflectance $r = \frac{I}{\pi F_{pivot}}$, where $\pi F_{pivot}$ is the solar irradiance at the scene at the pivot wavelength.

\subsubsection*{Voyager's ISS}

We also retrieve the data from the Jupiter encounter of the Voyager dataset on NASA's PDS \citep{Showalter_PDS_Voyager_archive}.
These images are already radiometrically calibrated and each pixel has an I/F value. Contrary to LORRI, the Voyager cameras do have color filters of various width and coverage. We decided to work with observations taken with the clear filter, which has the largest spectral coverage.

\section{Data processing}
This section details the different steps we take to use photometric data as accurate as possible. 

\subsection{Optical distortion}
The optical distortion of LORRI was predicted and confirmed in-flight to be under one pixel \citep{Noble_flightPerformanceLong_2009} so we do not use any additional correction. 

As for the Voyager dataset, reseau markings were embedded in the optics to measure the distortion and correct it postflight. We use the information made available by NASA's PDS Ring Node about the measured positions of the reseau in the images as well as their real position (GEOMA files) \citep{Showalter_PDS_Voyager_archive}. With a Delaunay triangulation and by computing the bilinear transformation of every triangle we correct for the distortion in the images. More details can be found in the supplementary material and in \citealt{Belgacem_ImageProcessingPrecise_PaSS_2019}.

\subsection{Geometric metadata correction}

Each image comes with a set of geometric metadata - i.e. information about the scene and the detector at the time it was taken. These elements suffer uncertainties and errors \citep{Jonniaux_RobustExtractionNavigation_GN&CS_2014} that lead to problems when projecting pixels to compute the geometry of the scene. 

We have developed several tools to correct these metadata. We use SurRender \citep{Brochard_ScientificImageRendering_IAC_2018}, an image renderer developed by Airbus Defence $\&$ Space, Toulouse to compute simulated images based on the metadata available and compare them to the real data. 

\begin{figure}[h!]
\centerline{\includegraphics[width=16cm]{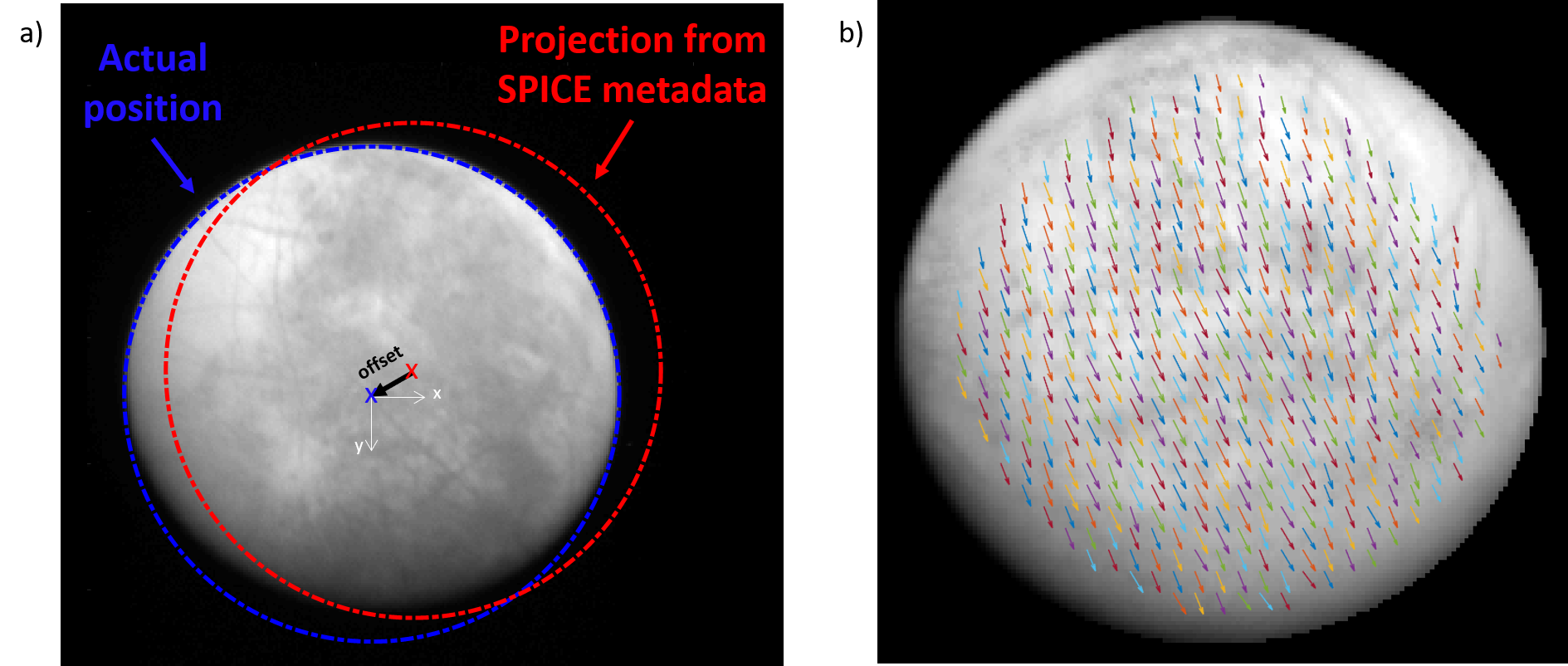}}
\caption{Illustration of the metadata errors observed in the dataset. a) Actual limb versus limb projected from the SPICE metadata b) Zoom on Europa showing the displacement vector field between actual image and the estimated attitude of Europa from the metadata.}
\label{fig:correct_pointing+att}
\end{figure}

We first correct for the pointing error by modifying the camera attitude using an optimization-based registration function. Fig. \ref{fig:correct_pointing+att}a illustrates the pointing error with an example of a LORRI observation. 

Second, we correct the attitude of the moon by computing the optical flow between simulated and real images (fig. \ref{fig:correct_pointing+att}b) and converting it to rotation using Kabsh algorithm \citep{Kabsch_DiscussionSolutionBest_ACSA_1978}. More information can be found in the supplementary material and in \citealt{Belgacem_ImageProcessingPrecise_PaSS_2019}.

On the Voyager data, correcting the spacecraft pointing and the moon's attitude is not enough. We also correct the camera position, and the corresponding Voyager distance to the moon. To do that, we use SurRender to compute simulations with varying distances between camera and moon and estimate the similarity to the real image with the Structural SIMilarity index (SSIM). The SSIM has notably been shown to improve more conventional techniques in object tracking in videos \citep{Loza_StructuralSimilarityBased_MVaA_2007}. The simulation with the greatest SSIM gives us the most probable distance the image was taken from.

\subsection{Projection}

\begin{figure}[h!]
\centerline{\includegraphics[width=10cm]{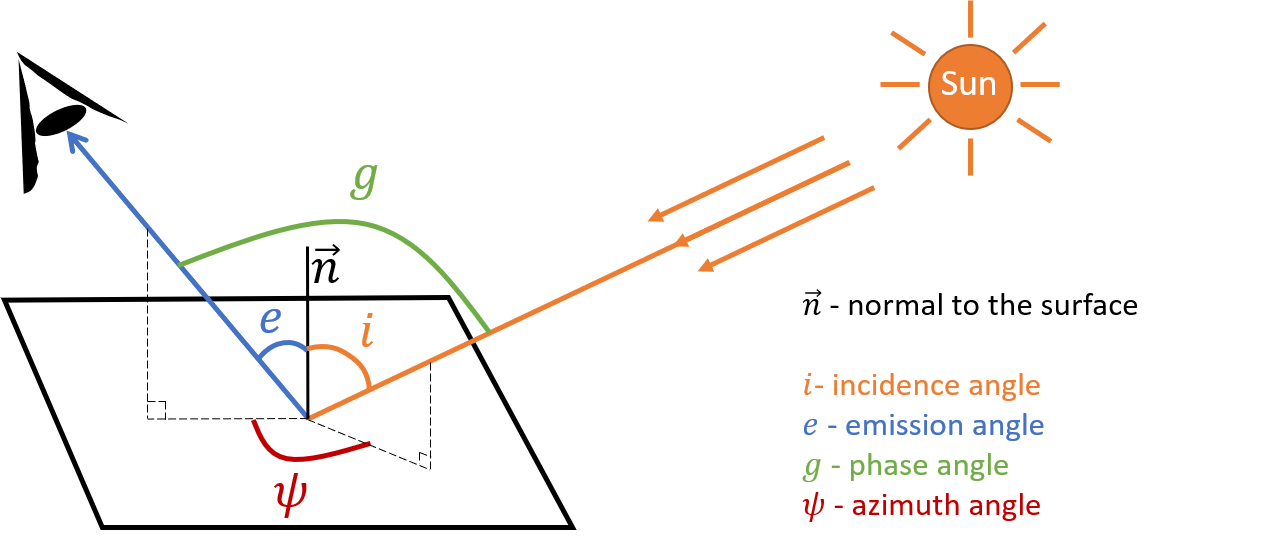}}
\caption{Definition of the viewing geometry angles}
\label{fig:geometry}
\end{figure}

After correction of the distortion and the metadata, we can safely use the information about the position and orientation of both the spacecraft and the moon to compute the geometry of observation i.e. the incidence (i), emission (e) and phase angle (g) (fig. \ref{fig:geometry}) as well as the latitude (lat) and longitude (lon) of the pixel center. 

We compute the geometry of each individual pixel by modeling Europa as an ellipsoid and use the radii information available in the SPICE kernels. More details can be found in the supplementary material. 

\subsection{Merging the datasets}
\begin{figure}[h!]
\centerline{\includegraphics[width=9cm]{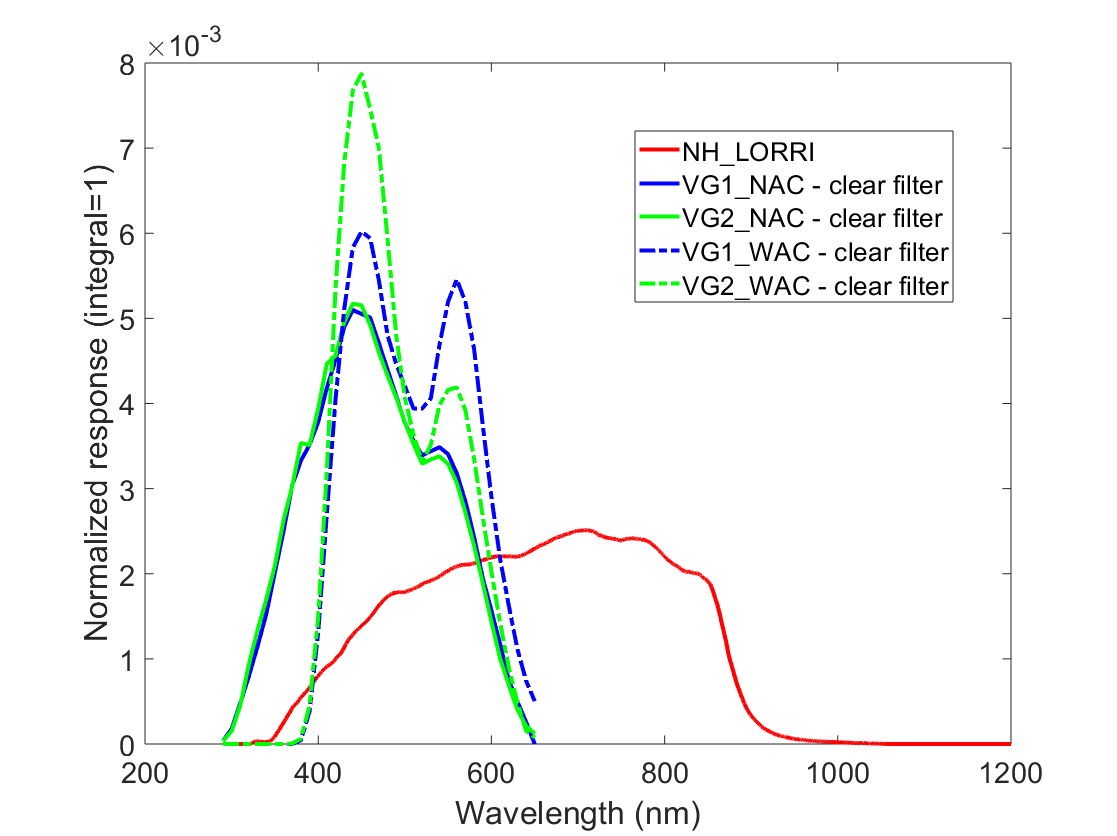}}
\caption{Comparison of LORRI'd instrument response and ISS' cameras response through the clear filter}
\label{fig:filters+rep}
\end{figure}

We combine both the LORRI and ISS datasets to maximize our surface coverage as well as the diversity of observation geometries (incidence, emission and phase angles) in this regional study. 
One major issue in doing that is that the color filters of LORRI and ISS are not matching as shown on fig. \ref{fig:filters+rep}. 

To merge all these observations, we make the choice of offsetting the Voyager data so it can be comparable to that of LORRI, taken as a reference. To do so, we have to choose a wavelength that is most representative of the entire filter. We choose the effective wavelength \citep{King_NoteConceptEffective_TAJ_1952} defined by:

\begin{equation}
\label{eq:def_eff}
\lambda_{eff} = \frac{\int P(\lambda)\lambda d\lambda}{\int P(\lambda) d\lambda}
\end{equation}

where $P(\lambda)$ is the instrument response.

\begin{figure}[h!]
\centerline{\includegraphics[width=10cm]{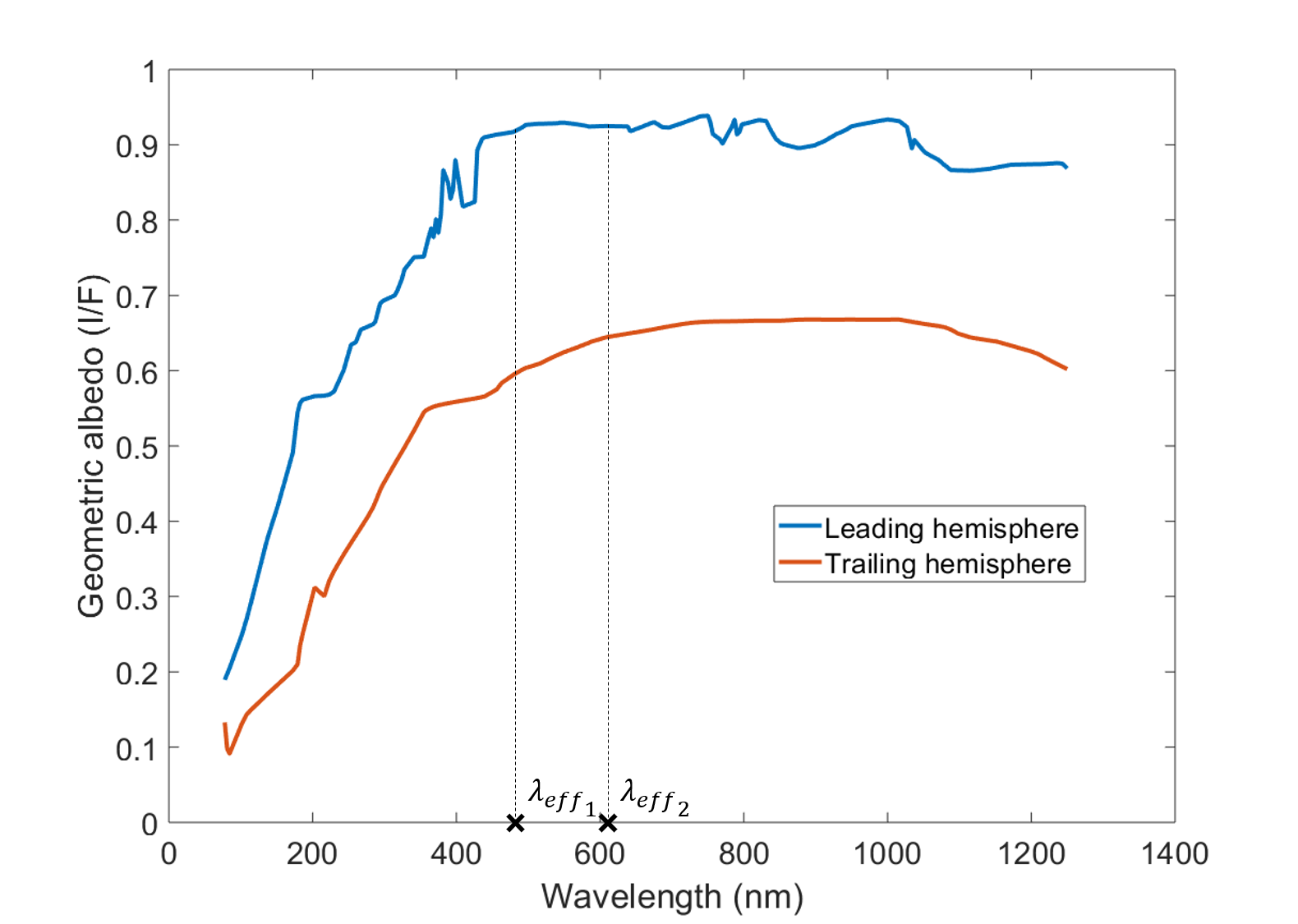}}
\caption{Europa spectra \citep{Calvin_SpectraIcyGalilean_JoGR_1995} where $\lambda_{eff1}$ is the effective wavelength of the VG1 NAC clear filter and $\lambda_{eff2}$ is the effective wavelength of NH LORRI}
\label{fig:europa_GA}
\end{figure}

We compute the effective wavelengths of the clear filters of every camera in the ISS as well as that of LORRI and summarize them in table \ref{tab:lambda_eff_clear}. Using Europa's spectrum \citep{Calvin_SpectraIcyGalilean_JoGR_1995} represented in fig. \ref{fig:europa_GA}, we can estimate that the ratio between Voyager and LORRI images should be equal to the ratio of the geometric albedo of Europa at the effective wavelengths. this ratio is computed for each camera and reported in table \ref{tab:lambda_eff_clear}. 

\begin{table}[h!]
	\centering
	\begin{tabular}{|l|c|c|c|c|c|}
		\hline
		  & NH\_LORRI & VG1\_NAC  & VG1\_WAC & VG2\_NAC & VG2\_WAC      \\ \hline
		$\lambda_{eff}$& 623.6 & 467.2   & 508.0   & 464.5   & 492.4  \\ \hline
		
		ratio to LORRI & 1.00 & 1.06 & 1.03 & 1.06 & 1.04   \\ \hline

	\end{tabular}
	\caption{Effective wavelengths of LORRI and Voyager cameras through the clear filter}
	\label{tab:lambda_eff_clear}
\end{table}

\section{Definition of the regions of interest}

In the interest of having comparable observations under different geometries, we choose to keep images that were taken between $1 \times 10^6$ and $4\times 10^6$ kms, which translates to a ground resolution at the subspacecraft point ($e=0\degree$) between about $10$ and $30$ kms.

In the entire datasets of New Horizons and Voyager images, 18 images taken with LORRI and 39 taken with ISS meet these requirements. 

Additionally, to avoid any complication due to extreme geometries and Hapke's non physical domain, we only keep observations with incidence and emission angles below $70\degree$. 

\begin{figure}[h!]
\centerline{\includegraphics[width=16cm]{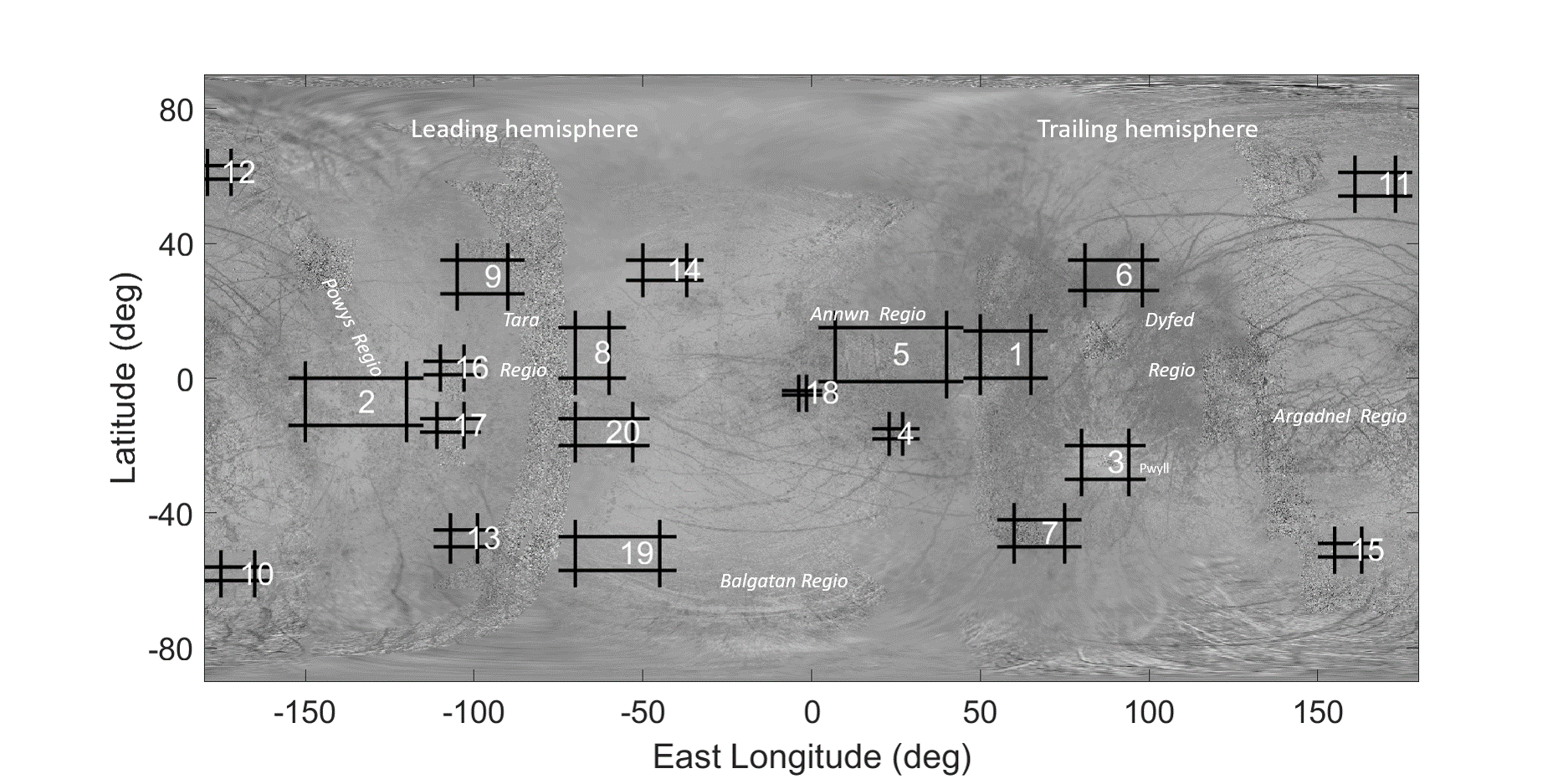}}
\caption{ROIs defined on color map by \citealt{Jonsson_MappingEuropa_2015}. The ROIs ID numbers are arbitrarily defined. The map projection is equidistant cylindrical, as all global maps of this article.}
\label{fig:map_zones}
\end{figure}

\begin{figure}[h!]
\centerline{\includegraphics[width=16cm]{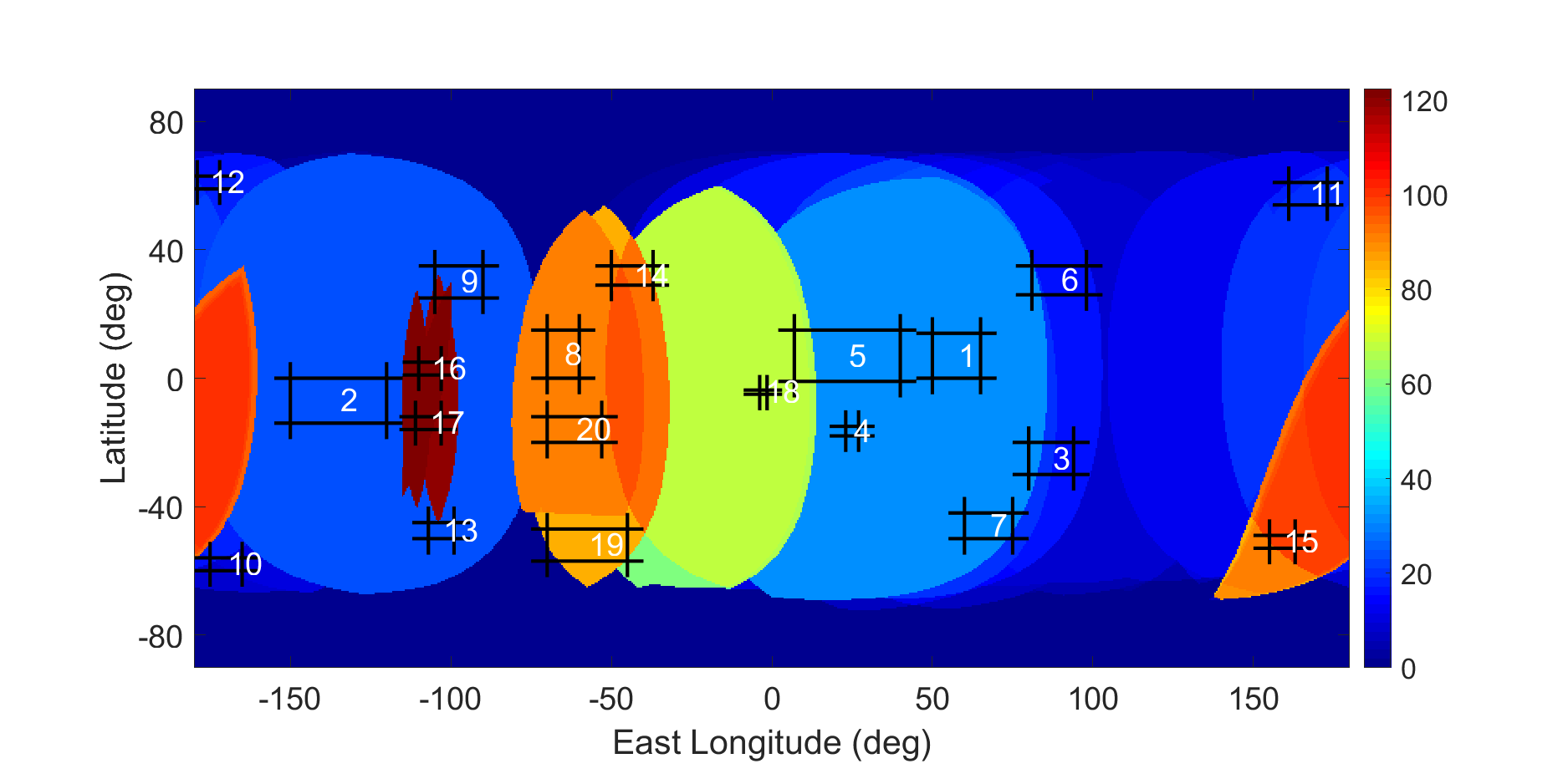}}
\caption{Phase coverage (dg) map of the 57 images from New Horizons' LORRI and Voyager's ISS. ROIs are plotted in black boxes.}
\label{fig:map_zones_ph}
\end{figure}

In these conditions, we try to maximize the phase coverage (dg = max(g)- min(g)) shown on fig. \ref{fig:map_zones_ph} and the diversity of geometries, as recommended by \citealt{Schmidt_EfficiencyBRDFSampling_I_2019}, \citealt{Schmidt_RealisticUncertaintiesHapke_I_2015} and \citealt{Souchon_ExperimentalStudyHapkes_I_2011}. We arbitrarily define $n_{ROIs}$=20 regions of interest (ROIs) on the surface of Europa in order to have good trade-off between size, number of images, number of pixels, phase coverage, albedo and geological terrains. We show these ROIs on fig. \ref{fig:map_zones} on a background color map from \citealt{Jonsson_MappingEuropa_2015}. Please note the map has been corrected and smoothed according to the process described in \citealt{Jonsson_MappingEuropa_2015} and therefore does not reflect a perfect ground truth.

The ROIs characteristics are available in table S2 in supplementary.

\section{Model}

We adopt the following notations:

\begin{itemize}
\item $i$: incidence angle
\item $e$: emission angle
\item $g$: phase angle
\item $\mu_{0} = \cos(i)$
\item $\mu= \cos(e)$
\end{itemize}

\subsubsection*{Hapke's photometric model}

Hapke model has been widely used to study the photometry of planetary surfaces in the Solar System. The model has been evolving ever since its first definition in the late 1980s. We will be focusing on the version described in \citealt{Hapke_TheoryReflectanceEmittance_book_1993}.
It is an empirical model based on the following equation:

\begin{equation}
\label{eq:def_hapke}
r = \frac{\omega}{4\pi}\frac{\mu_{0e}}{\mu_{0e} + \mu_{e}}\left[(1+B(g))P(g)+H(\mu_{0e})H(\mu_{e})-1\right]S(i, e, g)
\end{equation}

We note that the subscript $_e$ refers to the equivalent geometry in the rough case as defined by Hapke in \citealt{Hapke_TheoryReflectanceEmittance_book_1993}.

The model is described by six parameters:
\begin{itemize}
\item $\omega$: single scattering albedo, i.e. the fraction of interacting light scattered by a particle
\item $b$: asymmetry parameter, describes the isotropy of the scattering lobe
\item $c$: back-scattering fraction, describes the main direction of the scattering lobe
\item $\overline{\theta}$: macroscopic roughness
\item $h$: angular width of opposition surge
\item $B_0$: amplitude of opposition surge
\end{itemize}

The different functions of the model are described hereafter:

\begin{itemize}
\item $P(g)$: particle scattering phase function. Here we use a 2-term Henyey-Greenstein function (or HG2) \citep{Henyey_DiffuseRadiationGalaxy_TAJ_1941}:
\begin{equation}
\label{eq:def_HG2}
P(g) = (1-c)\frac{1-b^2}{\sqrt{(1+2bcos(g)+b^2)^3}} + c\frac{1-b^2}{\sqrt{(1-2bcos(g)+b^2)^3}}
\end{equation}
\item $H(x)$: multiple scattering function (improved expression from \citealt{Hapke_BidirectionalReflectanceSpectroscopy5_I_2002})
\begin{equation}
H(x) \approx \frac{1}{1-\omega \left[ r_0 + \frac{1-2r_0 x}{2} \ln\left( \frac{1+x}{x} \right) \right]}
\end{equation}
where $r_0=\frac{1-\gamma}{1+\gamma}$ and $\gamma=\sqrt{1-\omega}$.
\item $B(g)$: opposition effect function
\begin{equation}
B(g)=\frac{B_0}{1+\frac{1}{h}tan(\frac{g}{2})}
\end{equation}
\item $S(i, e, g)$: macroscopic roughness function (expression found in \citealt{Hapke_TheoryReflectanceEmittance_book_1993}
\end{itemize}

We will be working in REFlectance Factor (REFF) unit defined by: 
\begin{equation}
\label{eq:def_REFF}
REFF = r\frac{\pi}{\mu_0}
\end{equation}

\subsubsection*{Absolute reflectance}

\begin{figure}[h!]
\centerline{\includegraphics[width=13cm]{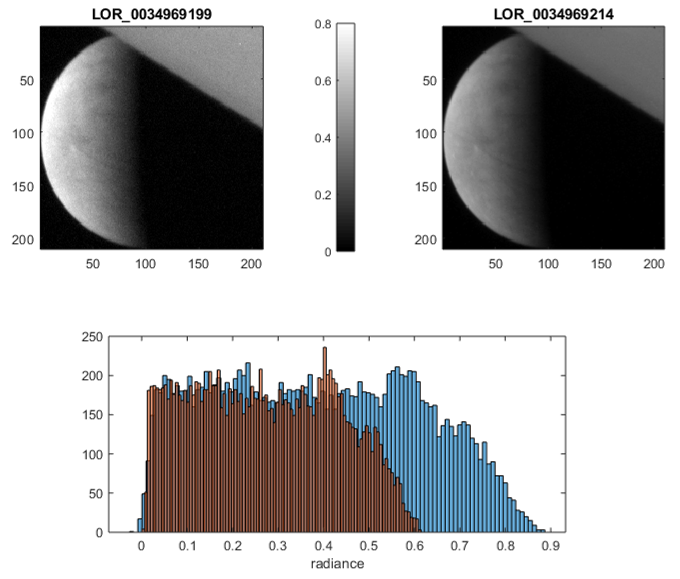}}
\caption{Comparison of two New Horizons observations taken a few minutes apart showing a significant error in absolute calibration. Top: the two LORRI images. Bottom: Histograms of the calibrated radiance of both images, only for Europa crescent (Jupiter has been masked).}
\label{fig:absolute_err}
\end{figure}

We have noticed various discrepancies in the dataset. For example, we have seen that two images of the same part of Europa taken in a similar geometry can show a difference in value of absolute radiance up to $35\%$. Fig. \ref{fig:absolute_err} illustrates that with two LORRI observations taken a few minutes apart. The histogram in the bottom part of the figure shows that even though they are very close in geometry, the absolute value of the radiance after calibration is significantly different. 

To account for these effects in the best way possible we have added image-dependent coefficients to modulate the value of the reflectance. Using that paradigm, the model we are trying to estimate is given by:

\begin{equation}
\label{eq:def_ref_alph}
r_{i,k} = (1+\alpha_k)\times r_{i,Hapke}
\end{equation}

Where:
\begin{itemize}
\item $r_{i,k}$ is the reconstructed reflectance of pixel $\#i$ in image $\#k$
\item $\alpha_k$ is the correction coefficient for image $\#k$
\item $r_{i,k,Hapke}$ is the synthetic reflectance given by the Hapke model (\ref{eq:def_REFF}) of pixel $\#i$ in image $\#k$
\end{itemize}

In the rest of this paper, we will be using the notations $m$, $m_i$ and $\alpha$.  $m$ is a generic denomination for all parameters. The $m_i$ vectors have a $6\times 1$ size and describe the Hapke parameters on ROI$\#$i, the $\alpha$ vector has a $1\times n_{imgs}$ size and describe the absolute calibration correction coefficients. The actual data will be noted $d$ and can be explicitly noted as all pixels, from all images: $\check{r_{i,k}}$. 

The goal of the inversion, is to estimate $m$ ($6 \times 20 + 57 = 176 $ unknowns), such as $\check{r_{i,k}}$ is close enough to $r_{i,k}$ ($5.0 \times 10^{4}$ known pixels).

\section{Inversion}

The 1980's 1990's have seen several studies on the photometry of Europa after we received data from the Voyager probes \citep{Buratti_VoyagerPhotometryEuropa_I_1983, Buratti_ApplicationRadiativeTransfer_I_1985,  Buratti_PhotometrySurfaceStructure_JoGR_1995, Domingue_EuropasPhaseCurve_I_1991,  Domingue_DiskResolvedPhotometric_I_1992}. If some of them used the Hapke model \citep{Buratti_PhotometrySurfaceStructure_JoGR_1995, Domingue_EuropasPhaseCurve_I_1991, Domingue_DiskResolvedPhotometric_I_1992}, they usually used a least-square minimization for the inverse problem. The main difficulty of such an approach would be that the minimization could easily converge towards a local minimum and not give the best possible solution. It is also difficult to have a realistic estimation of the uncertainty on the estimated parameters. 

That is why we opted for a Bayesian approach. This method has shown promise on Mars in a similar framework to estimate Hapke parameters with CRISM data \citep{Fernando_SurfaceReflectanceMars_JoGRP_2013, Schmidt_RealisticUncertaintiesHapke_I_2015, Fernando_MartianSurfaceMicrotexture_PaSS_2016} and has been revised since \citep{Schmidt_EfficiencyBRDFSampling_I_2019}. These studies have shown that a Bayesian framework is well suited for a photometric study and can give additional information on the uncertainties and constraints of the estimated parameters.

We adapted the method from \citealt{Schmidt_EfficiencyBRDFSampling_I_2019} to our problem. 

\subsection{Bayesian framework}

We use the theory of Inverse Problem based on the state of information described by a probability density function (PDF) \citep{Tarantola_InverseProblemTheory_book_2005}. When most common methods such as least square minimization try to find the best set of parameters to fit the data in a specific model, looking for the PDF gives a more complete understanding of the state of the solution. 

We introduce the notations:

\begin{itemize}
\item $\rho_M(m)$: the a priori information we have on the parameters i.e. mathematical intervals of definition etc.
\item $\sigma(m|d)$: the a posteriori probability density function i.e. the state of information on the parameters $m$ conditioned on the measurements $d$
\item $F$: the direct model we use to fit our data
\item $L$: the likelihood function defined by $\rho(d|m)$ which is the probability of obtaining the data $d$ conditioned on the parameters $m$
\end{itemize}

Using Bayes' theorem we can write;

\begin{equation}
\label{eq:PDF2proba}
\sigma_(m|d) = \frac{\rho(d|m)\rho_M(m)}{\rho_D(d)}
\end{equation}

We recognize the likelihood function $L(m)=\rho(d|m)$ which leads to:

\begin{equation}
\label{eq:PDF_apost}
\sigma_M(m|d) = k\rho_M(m)L(m,d)
\end{equation}

where:
\begin{itemize}
\item $k$ is a normalizing constant
\item $\rho_M$ is the a priori information on the parameters
\item $L$ is the likelihood function 
\end{itemize}

For a given ROI, as we are in the case of Gaussian additive noise in the data, the likelihood function can be expressed by:

\begin{equation}
\label{eq:def_L}
L = \frac{1}{\sqrt[N]{2\pi}\sqrt{|\Sigma|}}\exp\left(\frac{1}{2}((d-F(m))^T\Sigma^{-1}(d-F(m))\right)
\end{equation}

However, we want to do a global inversion and estimate simultaneously the Hapke parameters for each ROI as well as the correction coefficients for the absolute calibration of every image. Therefore, we define a global likelihood $\overline{L}$ as the product of the likelihood functions of each individual ROI:

\begin{equation}
\label{eq:def_newL}
\overline{L} = \prod_{k=1}^{n_{ROIs}} L_k
\end{equation}

\subsection{A priori information}
\subsubsection*{Parameters}

For each of the ROIs, we have 6 Hapke parameters to estimate. The only a priori information that we take into account for each of them is the physical definition: they must follow a uniform law on the $[0,1]$ interval except for $\overline{\theta}$ that evolves in the $[0,45\degree]$ interval. Additionally, the correction coefficients $\alpha_k$ are assumed to follow a centered Gaussian distribution with a standard deviation $\sigma_L=30\%$ in agreement with observed discrepancies (see fig. \ref{fig:absolute_err}).

\subsubsection*{Data}
We assume that the data is described by the model defined by eq. \ref{eq:def_ref_alph}. However, we know that there are still noise and uncertainties that are not accounted for. We assume that these uncertainties follow a centered Gaussian distribution with a standard deviation of $\sigma_N=30\%$. This value is fixed as a maximum boundary of the noise level in the image. To introduce this in the inversion process, we define the covariance matrix $\Sigma$, where $n_{pix}$ is the total number of pixels and $\sigma_{N,i}$ is the individual level of noise associated to pixel $\#i$:

\begin{equation}
\label{eq:def_diag_covarmat}
\bigsigma = 
\begin{array}{ccc}
\begin{pmatrix}
\sigma_{N,1}^2 & ~ & \Hugzero & ~ \\
~ & \sigma_{N,2}^2 &  ~&  ~\\
~ & ~ & \ddots & ~\\
~ &  \Hugzero & ~ & \sigma_{N,n_{pix}}^2\\
\end{pmatrix}
\end{array}
\end{equation}

We will see in section \ref{sec:results} that the actual level of noise present in the estimated model fits is much lower than that.

\subsection{A posteriori PDF sampling method}
Statistical methods have been developed to efficiently sample a posteriori probability distributions. We choose a Markov Chain Monte Carlo (MCMC) sampling method based on a Metropolis-Hastings algorithm \citep{Mosegaard_MonteCarloSampling_JoGRSE_1995} with an efficient sampling algorithm \citealt{Schmidt_EfficiencyBRDFSampling_I_2019}, adapted to this particular problem. The detailed algorithm is in the supplementary material.

\subsection*{Estimators}
Once the a posteriori PDF is sampled, we consider two estimators to summarize the solution on $m$ and $\alpha$:
   
\begin{itemize}
\item the mean a posteriori defined by the integral of the a posteriori PDF and can be completed by the standard deviation of the a posteriori distribution
\item the maximum likelihood defined by the set of parameters that maximize the likelihood function $\overline{L}$.
\end{itemize}

\section{Results}
\label{sec:results}
We have realized a global estimation by determining simultaneously the best Hapke parameters for all the ROIs as well as the absolute calibration correction coefficients for all the images. 

We used a computing system composed of 2 Intel(R) Xenon(R) CPUs E5-2690 v4 @2.6Ghz with 28 cores each. We realized a simulation over a total of 25 days 7 hours 30 minutes, reaching a Markov chain length of $4.05\times 10^7$ iterations. It is worth noting that since the algorithm is mostly iterative, the 56 cores were only used during the direct model computation. A burn-in period is expected at the beginning of the chain before converging to the solution. We optimize this burn-in period empirically and settle on a value of $2.0\times 10^7$ iterations. This leaves a chain of $2.05\times 10^7$ iterations to describe the solution. Once the convergence has been established, any additional chain from this point forward can be considered statistically equivalent to the initial Markov chain and all chains of iterations can be stacked virtually extending it. In the interest of computer efficiency, we ran several chains on different servers after the convergence and we extended the initial Markov chain to $10.83\times 10^7$ iterations, burn-in period excluded. This would have taken 67 days 16 hours and 30 minutes if we only had one chain on our fastest server. The acceptation rate of the chain is very low at $0.01\%$ which is to be expected with that many parameters to estimate simultaneously and that many data points and justifies the need for such a long computing process.

Before describing the results for each parameter, we assess how well the model fits the data with the estimated parameters. The Root Mean Square Deviation $RMSD$ is defined by the following equation, where $\textbf{d}$ is the data and $\textbf{d}_{fit}$ is the model fit:

\begin{equation}
\label{eq:def_RMSE}
RMSD = \sqrt{\frac{(\textbf{d}-\textbf{d}_{fit})^2}{N}}
\end{equation}

\begin{figure}[h!]
\centerline{\includegraphics[width=11cm]{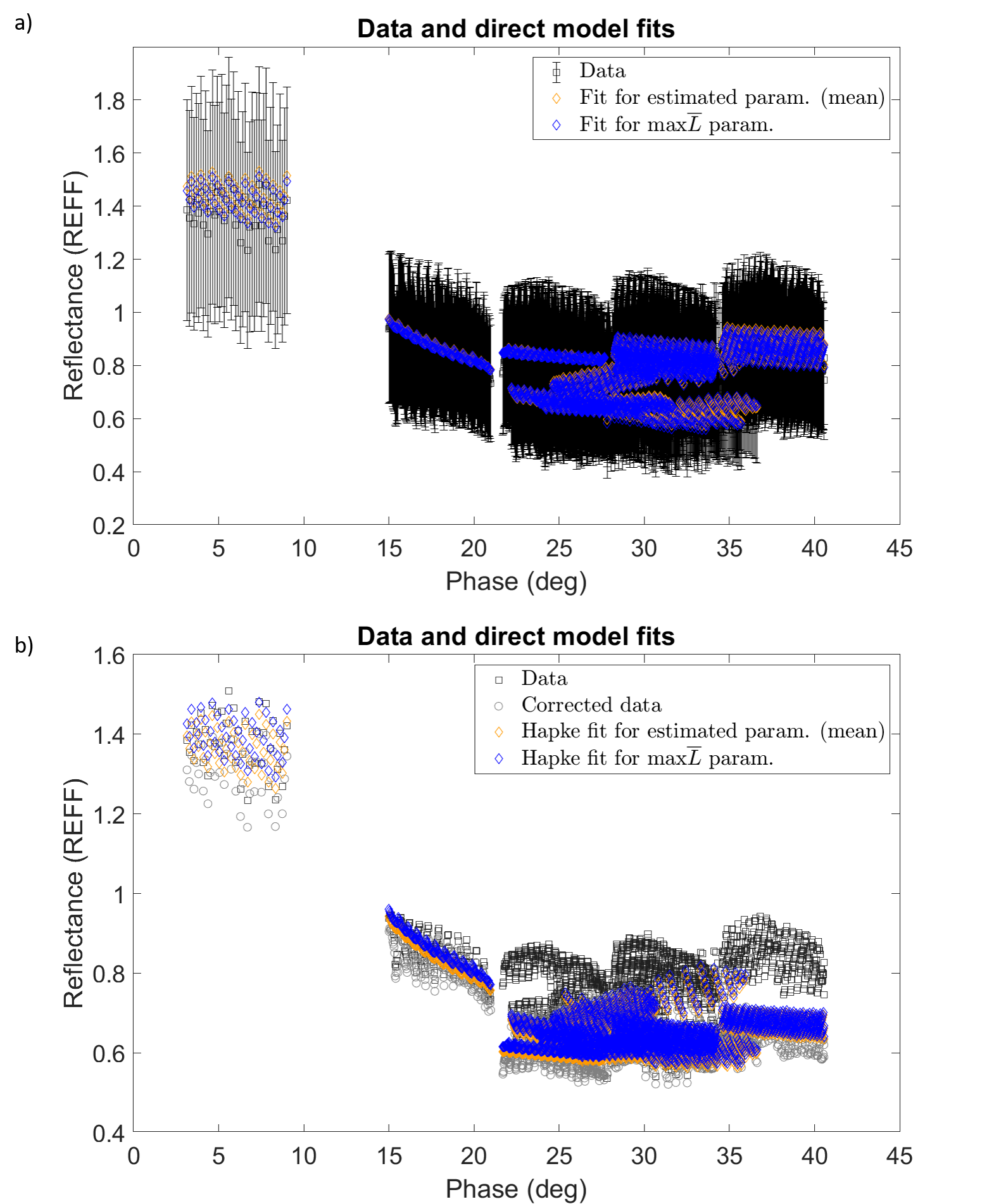}}
\caption{Data and best estimated model for ROI$\#$3 on Europa a) comparison of the reflectance data and the direct model from eq. \ref{eq:def_ref_alph}. The error bars represent 2$\sigma_N$ and the data are artificially spread out in phase to see the different pixels at the same phase angle. b) comparison of the data, the corrected data (with our estimated absolute level correction coefficients) and the estimated Hapke model}
\label{fig:fit_ex}
\end{figure}

The last column of table \ref{tab:results} lists the $RMSD$ for all ROIs, with $d_{fit}$ generated with the mean estimator. In percentage of the mean REFF value of the ROIs, this represents a $0.2\%$ (ROI$\#$10) to $4.7\%$ (ROI$\#$5) divergence from the actual data with a mean value of $1.3\%$. We have to note that this is much less that the $30\%$ imposed as a prior noise level. Fig. \ref{fig:fit_ex} shows the data and fits generated with the estimated parameters both with the complete direct model (eq. \ref{eq:def_ref_alph}) and after correction of the absolute level in the images.

\subsection{Absolute level corrections}

\begin{figure}[h!]
\centerline{\includegraphics[width=18cm]{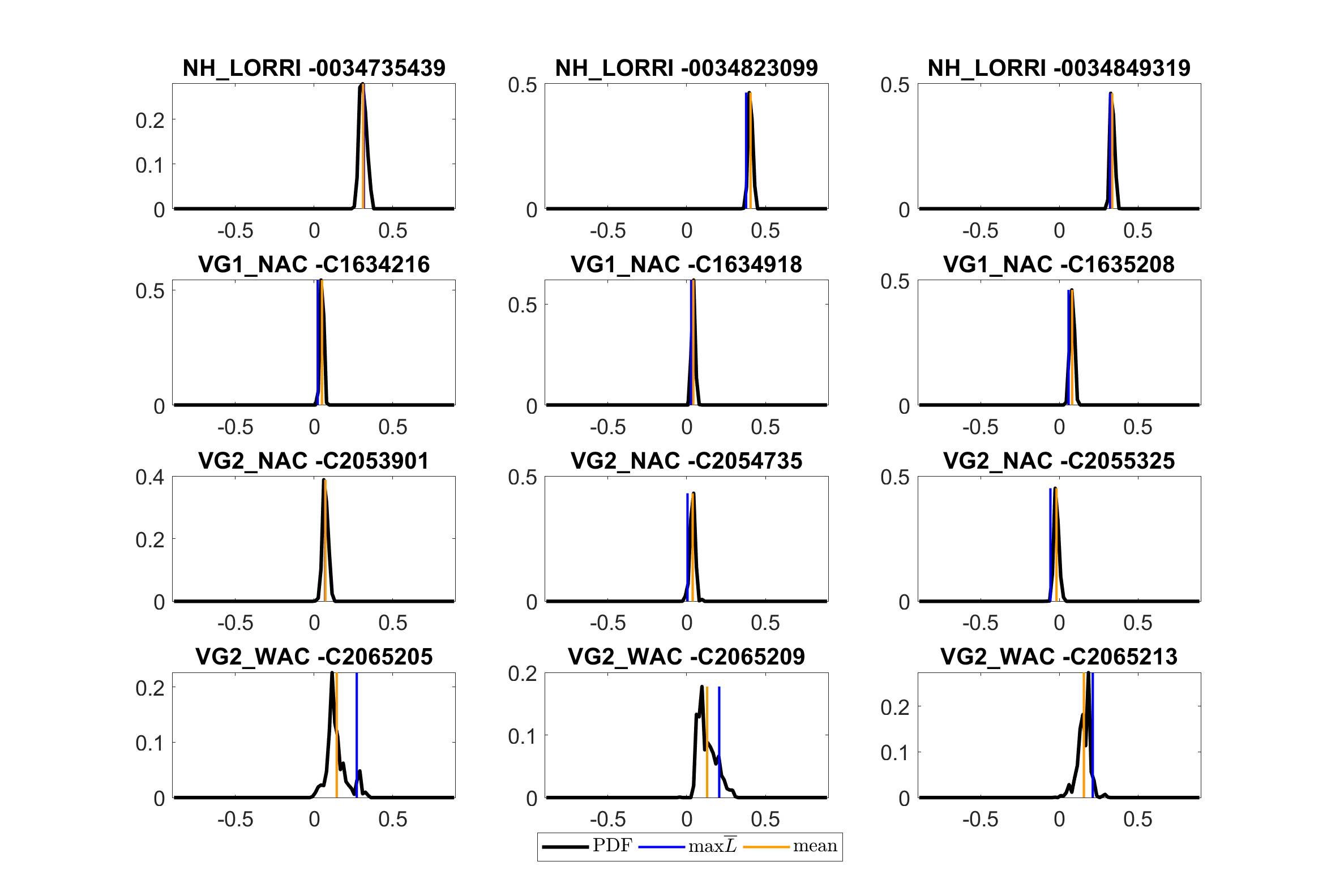}}
\caption{Probability Density Function (PDF) for the correction coefficients $\alpha$ of a selection of 12 representative images.}
\label{fig:PDF_alpha}
\end{figure}

Fig.\ref{fig:PDF_alpha} shows the a posteriori Probability Density Functions (PDFs) for the correction coefficients of a selection of images. We can first note that all of them are very well constrained since PDFs are much narrower than the initial Gaussian distribution with a $30\%$ standard deviation.

\begin{figure}[h!]
\centerline{\includegraphics[width=11cm]{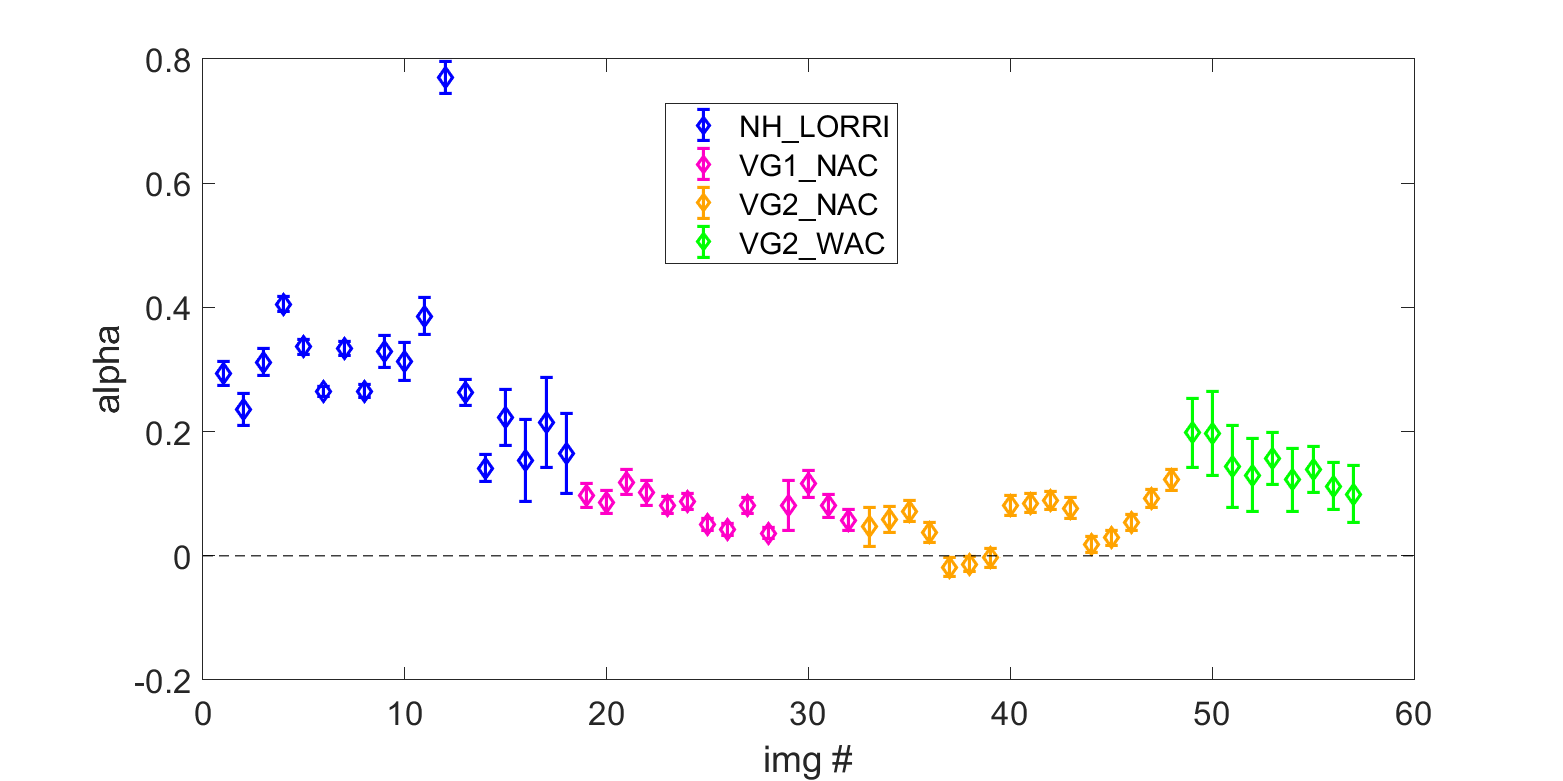}}
\caption{Mean and standard deviation of the PDF of correction coefficients $\alpha$ for all the images used in the inversion}
\label{fig:alpha}
\end{figure}

Fig. \ref{fig:alpha} shows all the values of the coefficients for the entire collection of images we used. Most values are very close to zero, especially for the Voyager images. We can also note a positive bias on most images, notably the ones taken by Voyager 2 wide angle camera (in green in fig. \ref{fig:alpha}). New Horizons' LORRI images present the most scattered and the greatest correction coefficients $\alpha$, in agreement with what we have illustrated in fig. \ref{fig:absolute_err}. The origin of these discrepancies are yet unknown.

\begin{figure}[h!]
\centerline{\includegraphics[width=13cm]{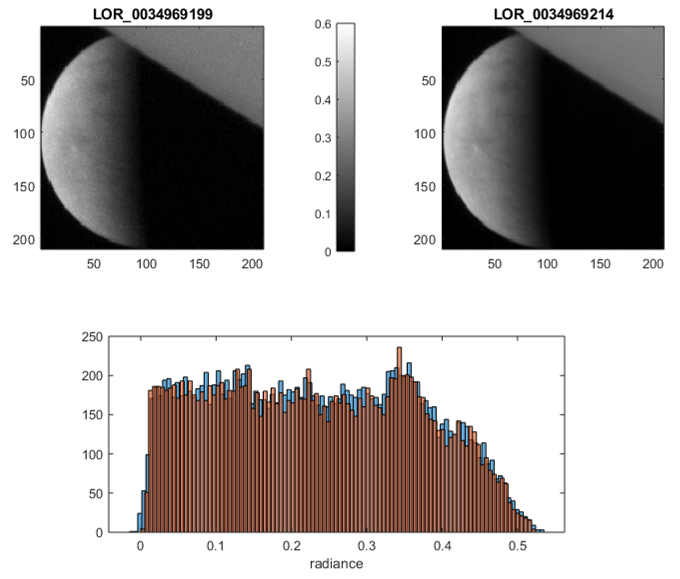}}
\caption{Same as fig. \ref{fig:absolute_err} but after level correction with estimated coefficients $\alpha$.}
\label{fig:absolute_err_corrected}
\end{figure}

Moreover, fig. \ref{fig:absolute_err_corrected} shows how effective our strategy to correct these discrepancies is by showing the same images as fig. \ref{fig:absolute_err} after correction.

\subsection{Analysis of the estimated Hapke parameters}

\begin{figure}[h!]
\centerline{\includegraphics[width=18cm]{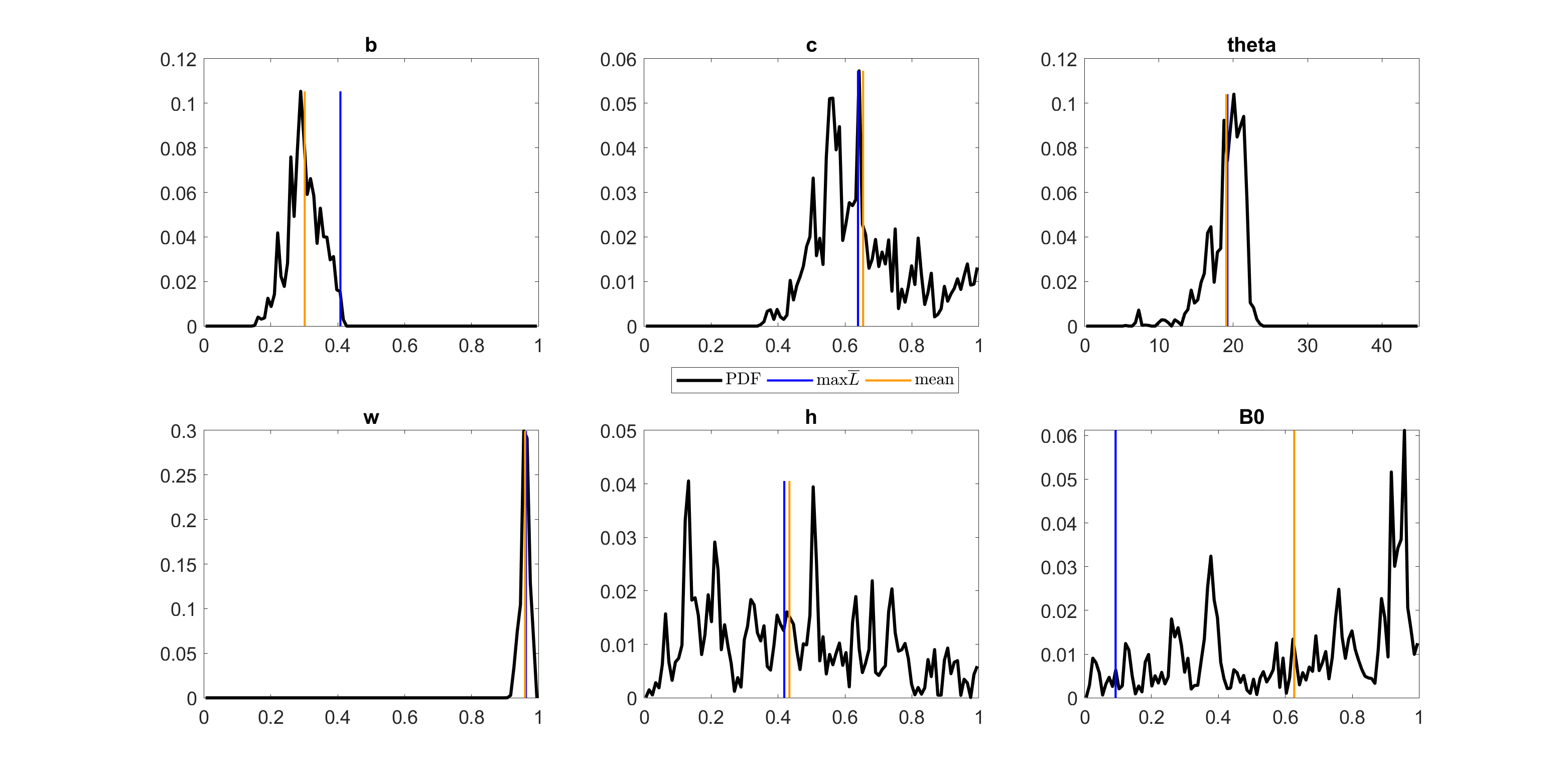}}
\caption{Probability Density Functions for the six Hapke parameters on ROI$\#$16. Color lines represent the global maximum likelihood $\overline{L}$ in blue) and the mean of the PDF (in orange).}
\label{fig:PDF}
\end{figure}

Fig. \ref{fig:PDF} shows an example of a posteriori PDFs for ROI$\#16$ as well as the mean value and the maximum global likelihood $\overline{L}$.

We use two main criteria to filter out non relevant results on each ROI:
\begin{itemize}
\item bimodality of the single scattering albedo $\omega$: as $\omega$ is usually the easiest photometric parameter to constrain, it is our view that a bimodal $\omega$ distribution over a particular ROI would most probably be the sign of an under-constrained solution due to a lack of geometrical diversity \citep{Fernando_SurfaceReflectanceMars_JoGRP_2013, Schmidt_RealisticUncertaintiesHapke_I_2015}. It could be argued that a bimodal PDF would describe two distinct photometric behaviors in the ROI but work by \citealt{Schmidt_EfficiencyBRDFSampling_I_2019} showed that the estimated photometry of a mixing of materials with different properties would not result in a bimodal PDF for $\omega$ but rather in a single value that had no easily interpretable physical significance .
\item non-uniformity criterion: we can consider that a solution has been found if the a posteriori PDF is different enough from the a priori PDF i.e. if the a posteriori PDF is non-uniform. We used $k>0.5$, as defined in \citealt{Fernando_SurfaceReflectanceMars_JoGRP_2013}, to test if the distribution is non-uniform. We only keep results for which the PDFs of at least $\omega$, $b$, $c$ and $\overline{\theta}$ are non-uniform.
\end{itemize}

All ROIs passed these tests, as shown in table S3 in the supplementary. For each parameter, the mean and standard deviation of the distribution gives an estimation of the solution and its uncertainties. Real PDFs may be different from Gaussian, and thus difficult to fully describe with 2 parameters only (see fig. \ref{fig:PDF}), but we think that this strategy is a good trade-off between complexity and accuracy.

Table \ref{tab:results} gives the extensive results over the ROIs. We only show the parameters $\omega$, $b$, $c$ and $\overline{\theta}$. The opposition effect and the parameters $h$ and $B_0$ will be discussed in a subsequent section.

\begin{table}[h!]
	\centering
\begin{tabular}{|l|c|c|c|c|c|c|c|c|c|c|c|c|c|}%
	\hline
    ~ & \multicolumn{2}{|c}{$\omega$}
    & \multicolumn{2}{|c}{$b$}& \multicolumn{2}{|c}{$c$} 
	& \multicolumn{2}{|c}{$\overline{\theta}$}
	& ~
	
	\\
	
	ROI & mean & std  
	& mean & std & mean & std 
	& mean & std 
	& RMSD 

    \begin{filecontents*}{params_invhapke_parallel70_sigL=30_tol=30.csv}
ROI,bmean,bstd,beu,cmean,cstd,ceu,tmean,tstd,teu,wmean,wstd,weu,hmean,hstd,heu,Bmean,Bstd,Beu,RMSD,ok,yes,no
1,0.46,0.02,1,0.17,0.02,1,15.88,5.84,1,0.92,0.01,1,0.1,0.07,1,0.75,0.13,0.98,0.015,ok,yes,no
2,0.43,0.02,1,0.46,0.04,1,13.03,5.08,1,0.95,0.01,1,0.6,0.25,0.56,0.2,0.09,0.99,0.019,ok,yes,no
3,0.32,0.02,1,0.83,0.07,1,23.27,4.18,1,0.91,0.02,1,0.59,0.23,0.54,0.44,0.12,1.03,0.014,ok,yes,no
4,0.42,0.05,1,0.31,0.08,0.99,18.24,9.45,1,0.9,0.02,1,0.5,0.32,0.73,0.33,0.15,1.05,0.005,ok,yes,no
5,0.35,0.02,1,0.35,0.03,1,11.95,5.03,1,0.89,0.01,1,0.16,0.05,1,0.89,0.07,1.02,0.025,ok,yes,no
6,0.31,0.02,1,0.9,0.05,1,11.33,5.58,1,0.7,0.02,1,0.77,0.19,0.98,0.34,0.1,1.03,0.01,ok,yes,no
7,0.3,0.04,1,0.48,0.08,1,26.65,7.3,1,0.93,0.02,1,0.59,0.27,0.18,0.78,0.13,1.03,0.011,ok,yes,no
8,0.43,0.04,1,0.48,0.08,1,7.23,4.07,1,0.95,0.01,1,0.47,0.3,0.68,0.23,0.12,1.01,0.007,ok,yes,no
9,0.5,0.04,1,0.2,0.06,1,23.05,1.49,1,0.99,0.01,1,0.45,0.26,0.4,0.48,0.21,1.04,0.006,ok,yes,no
10,0.26,0.06,1,0.83,0.12,1,17.67,11.9,1,0.95,0.03,1,0.42,0.28,0.76,0.47,0.28,0.24,0.002,ok,yes,no
11,0.43,0.07,1.01,0.37,0.08,1,16.22,9.8,1,0.98,0.01,1,0.42,0.27,0.44,0.44,0.24,0.59,0.004,ok,yes,no
12,0.23,0.14,0.95,0.67,0.21,0.71,17.49,11.42,1,0.95,0.05,1,0.6,0.24,0.68,0.66,0.2,1.02,0.005,ok,yes,no
13,0.27,0.09,1.02,0.61,0.16,1.28,16.79,11.33,1,0.98,0.02,1,0.39,0.3,1.01,0.69,0.25,0.8,0.003,ok,yes,no
14,0.36,0.03,1,0.68,0.08,1,5.93,4.16,1,0.95,0.01,1,0.3,0.19,1.06,0.37,0.1,1.02,0.009,ok,yes,no
15,0.36,0.04,1,0.84,0.11,0.99,8.25,4.27,1,0.88,0.01,1,0.25,0.26,1.49,0.41,0.22,0.76,0.007,ok,yes,no
16,0.3,0.05,1,0.65,0.15,0.99,19.06,2.5,1,0.96,0.01,1,0.43,0.25,0.58,0.63,0.29,0.65,0.004,ok,yes,no
17,0.25,0.04,1,0.74,0.11,0.99,18.91,1.77,1,0.94,0.01,1,0.73,0.18,1,0.83,0.12,0.99,0.004,ok,yes,no
18,0.26,0.08,1.01,0.78,0.22,1.16,21.13,11.65,1,0.88,0.04,1,0.41,0.21,0.86,0.47,0.22,0.79,0.003,ok,yes,no
19,0.6,0.02,1,0.22,0.04,1,15.37,1.55,1,0.99,0.01,1,0.23,0.24,1.02,0.51,0.33,1.17,0.015,ok,yes,no
20,0.28,0.03,1,0.85,0.08,1,13.24,4.76,1,0.94,0.01,1,0.51,0.27,0.22,0.23,0.11,1,0.009,ok,yes,no
\end{filecontents*}
\csvreader[head to column names]{params_invhapke_parallel70_sigL=30_tol=30.csv}{}
    {\\ \hline \ROI & \wmean & \wstd 
    & \bmean & \bstd  & \cmean & \cstd 
    & \tmean & \tstd 
    & \RMSD
    }
    \\ \hline
\end{tabular}
	\caption{Value of the estimated Hapke parameters over all the ROIs}
	\label{tab:results}
\end{table}

\newgeometry{top=1cm,bottom=3.5cm}
\begin{figure}[h!]
\centering
	\begin{subfigure}[h!]{0.71\textwidth}
	\includegraphics[width=\textwidth, trim={0cm 1cm 0cm 0.5cm}, clip=true]{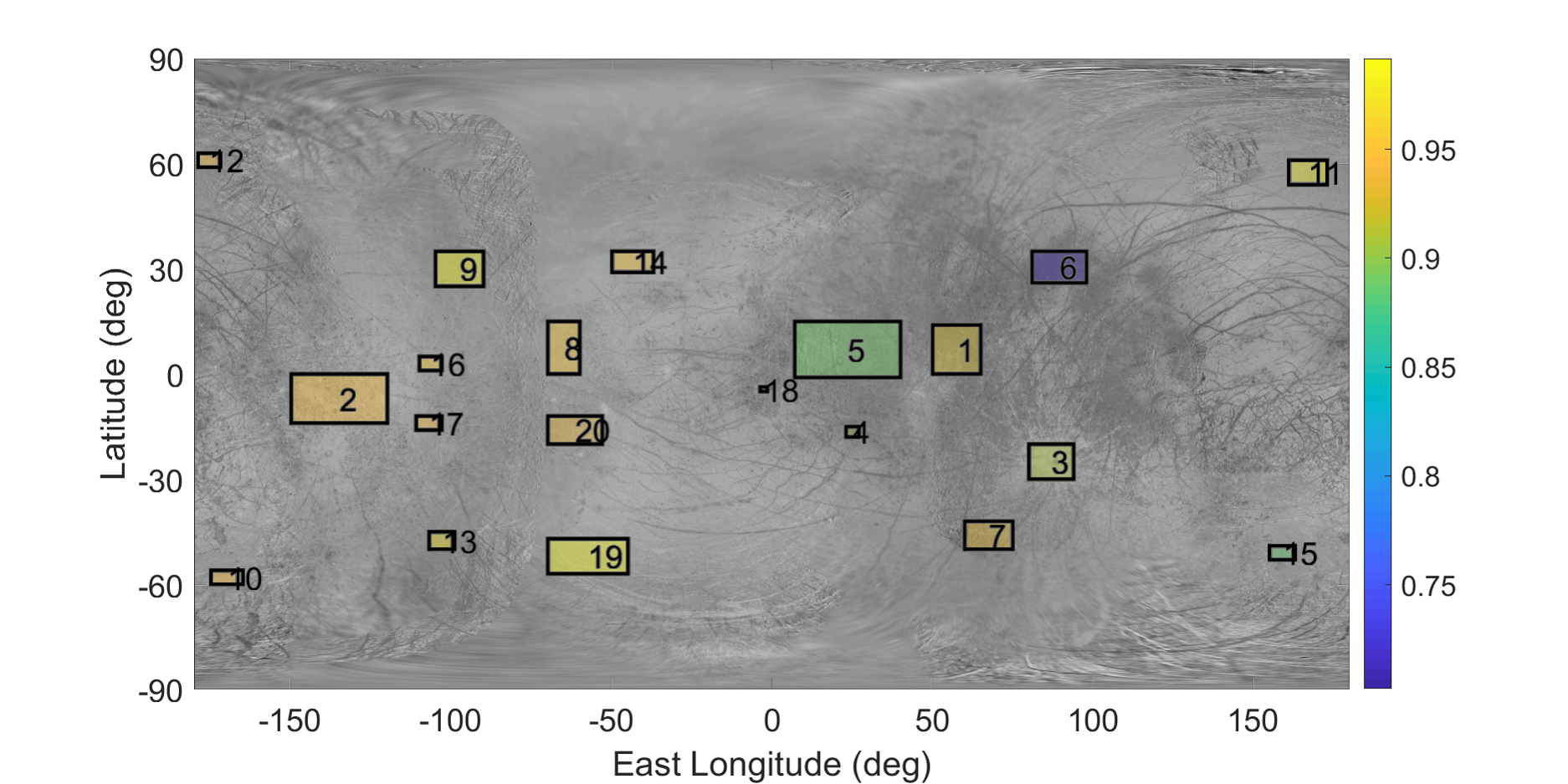}
	\subcaption{Values of the single scattering albedo $\omega$}
	\label{fig:eur_map_omega}
	\end{subfigure}

	\begin{subfigure}[h!]{0.71\textwidth}
	\includegraphics[width=\textwidth, trim={0cm 1cm 0cm 0.5cm}, clip=true]{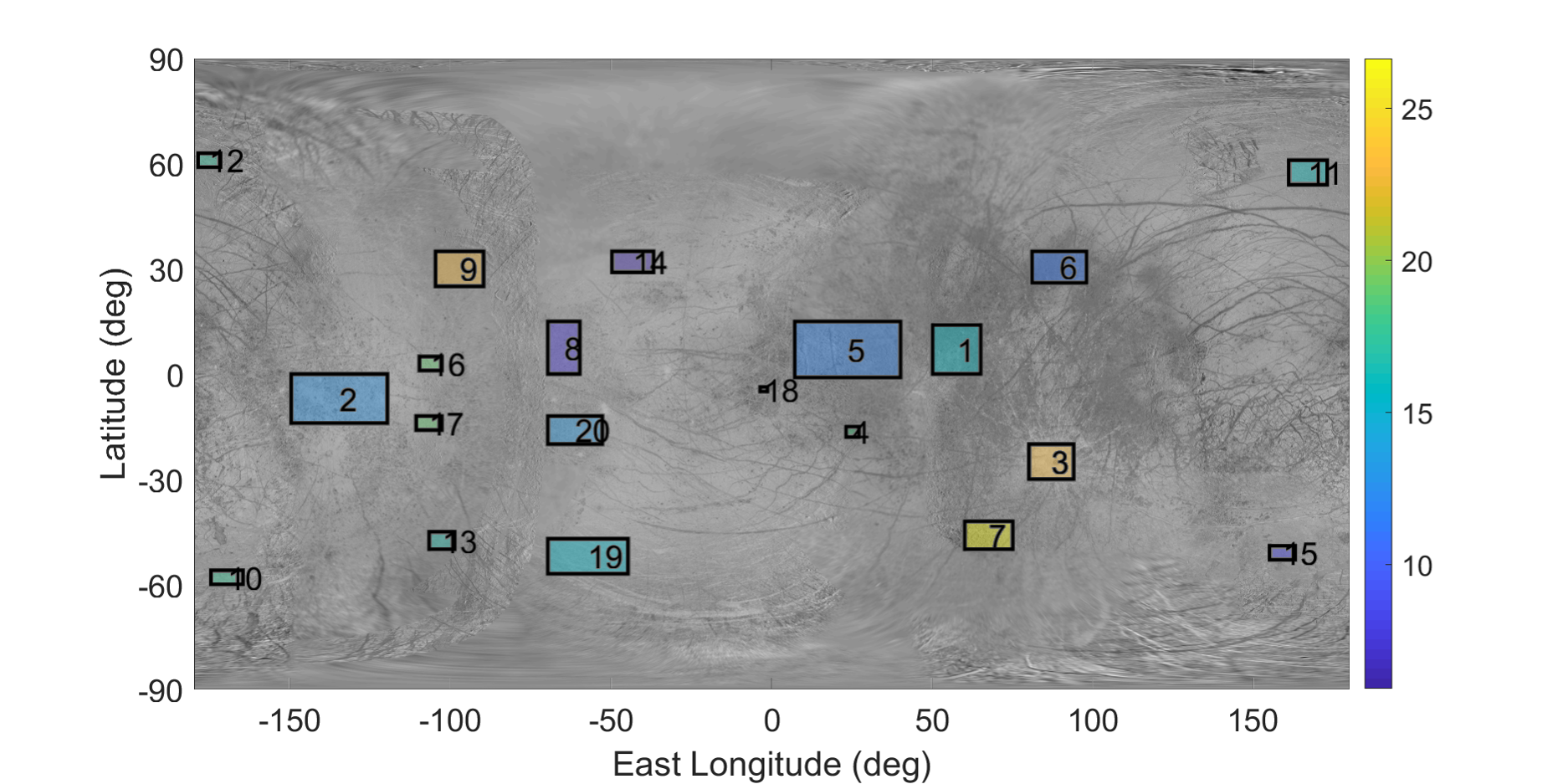}
	\subcaption{Values of the  macroscopic roughness $\overline{\theta}$}
	\label{fig:eur_map_theta}
	\end{subfigure}
	
	\begin{subfigure}[h!]{0.71\textwidth}
	\includegraphics[width=\textwidth, trim={0cm, 1cm, 0cm, 0.5cm}, clip=true]{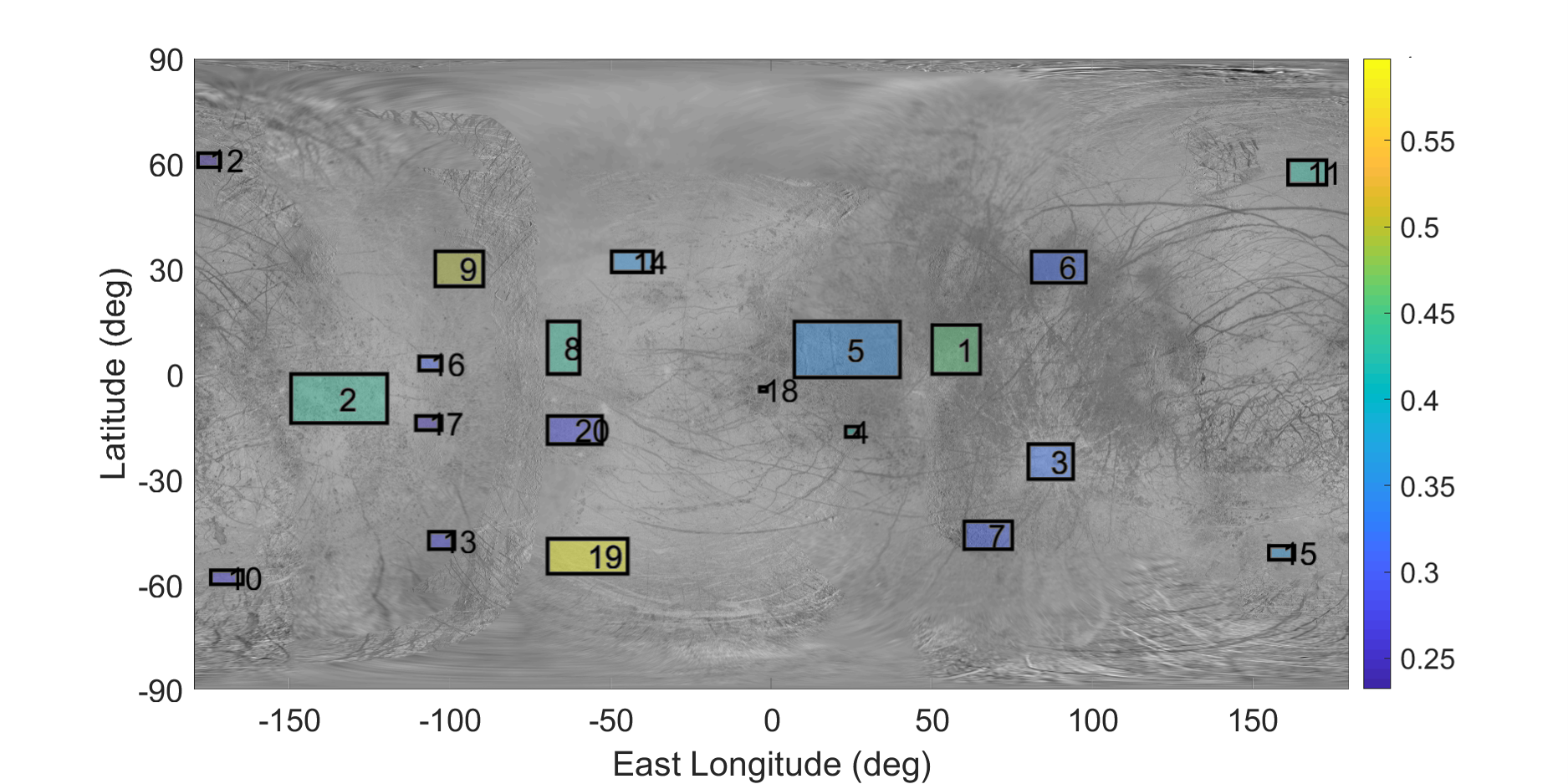}
	\subcaption{Values of the asymmetry parameter $b$}
	\label{fig:eur_map_b}
	\end{subfigure}

	\begin{subfigure}[h!]{0.71\textwidth}
	\includegraphics[width=\textwidth, trim={0cm, 0cm, 0cm, 0.5cm}, clip=true]{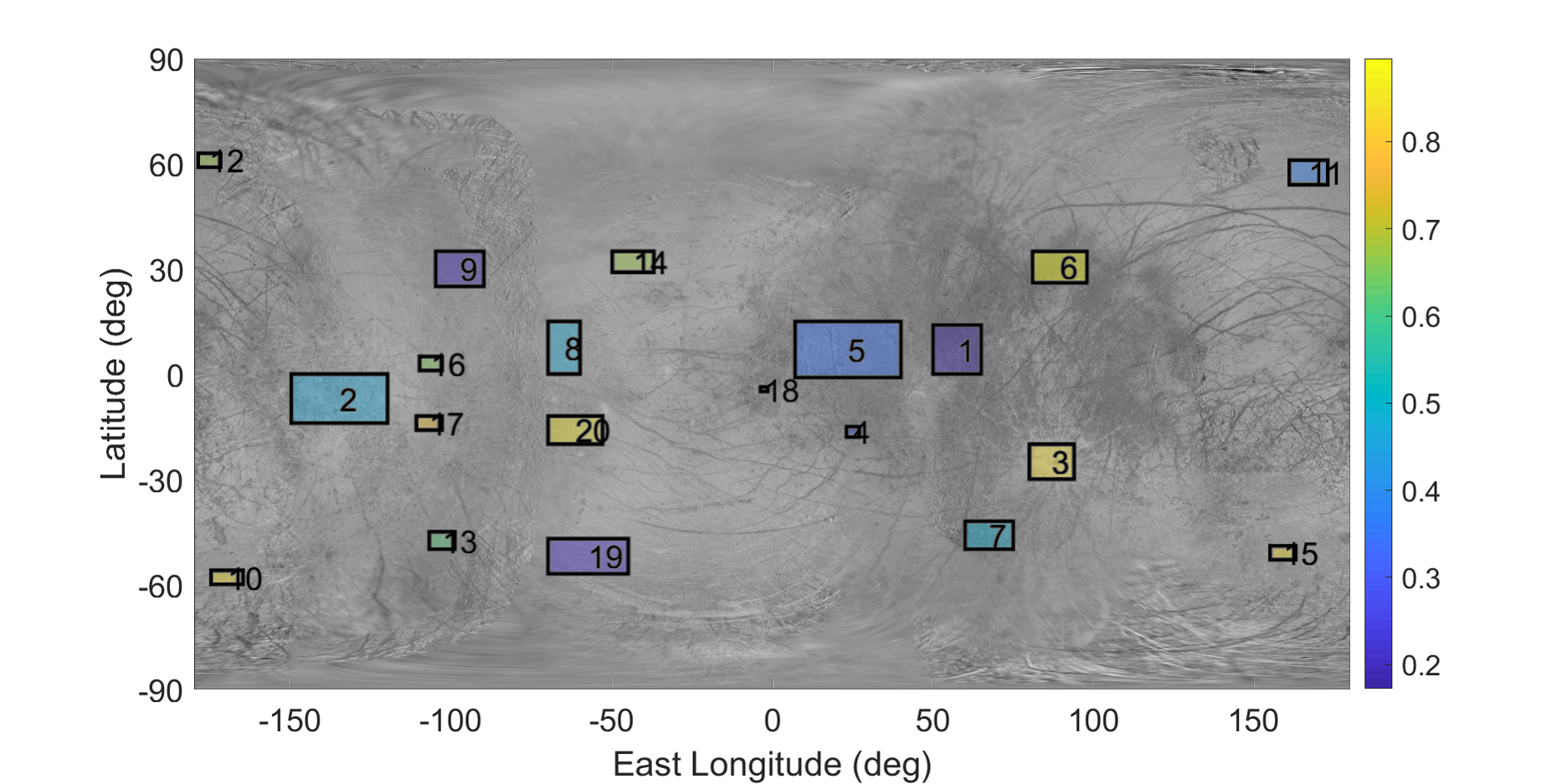}
	\subcaption{Values of the backscattering fraction $c$}
	\label{fig:eur_map_c}
	\end{subfigure}
	
	\caption{Value of the estimated Hapke parameters across the surface of Europa (mean of a posteriori PDF)}
	\label{fig:eur_hapke_maps}
\end{figure}

\restoregeometry

\subsubsection*{Single scattering albedo}
The single scattering albedo is the fraction of interacting light that is diffused by a particle. The higher $\omega$, the more diffusion over absorption. This behavior is directly linked to the particles properties such as optical constant, shape and internal structure. In the a posteriori PDFs, $omega$ is by far the most constrained parameter across Europa's surface confirming that it is the easiest parameter to estimate.

The single scattering albedo for all ROI is rather high across Europa's surface with an average of 0.93. One ROI ($\#6$) present a notably lower value of about $0.70$. It has been established that the trailing hemisphere (east longitudes $0\degree$ to $180\degree$) is darker than the leading (east longitudes $-180\degree$ to $0\degree$) \citep{Domingue_EuropasPhaseCurve_I_1991}. We verify this property in our results as the trailing hemisphere has an average single scattering albedo of 0.89 when the leading has an average of 0.95 (see fig. \ref{fig:eur_map_omega}). Even though single scattering albedo retrievals show some diversity, it is the least heterogeneous photometric parameter across the surface. The coherence between our retrievals and the albedo map (background of Fig. \ref{fig:eur_map_omega}) is not perfect. We interpret that this may due to the fact that J\'{o}nsson used images renormalization in the scope of reducing all intersection discrepancies in order to have the best eye-looking map. This process should preserve local details but may have led to some errors at large scale due to photometry.

\subsubsection*{Macroscopic roughness $\overline{\theta}$}

We find in our study very different values of the macroscopic roughness across the surface of Europa. They range from $5.9\degree$ to $26.6\degree$ and the mean value is $16.0\degree$. This is a slightly higher that the disk-averaged value found by \citealt{Domingue_EuropasPhaseCurve_I_1991} but values up to 15$\degree$ were found in the disk-resolved study of \citealt{Domingue_DiskResolvedPhotometric_I_1992}. These values are much higher than other icy bodies such as Enceladus for which \citealt{Verbiscer_PhotometricStudyEnceladus_I_1994} and  \citealt{Verbiscer_OppositionSurgeEnceladus_I_2005} both found consistent values around $7\degree$. However, values for Saturn's satellites Rhea and Phoebe reach $33\degree$ \citep{Ciarniello_HapkeModelingRhea_I_2011}.

$\overline{\theta}$ can very much depend on the way the opposition surge is parameterized \citep{Shepard_TestHapkePhotometric_JoGR_2007}. As we do not have a lot of data below 10$\degree$ and none below 1$\degree$ where the effects of the CBOE are strongest \citep{Helfenstein_GalileoObservationsEuropas_I_1998}, we can suppose that this has little incidence in our situation. 

\begin{figure}[h!]
\centerline{\includegraphics[width=18cm]{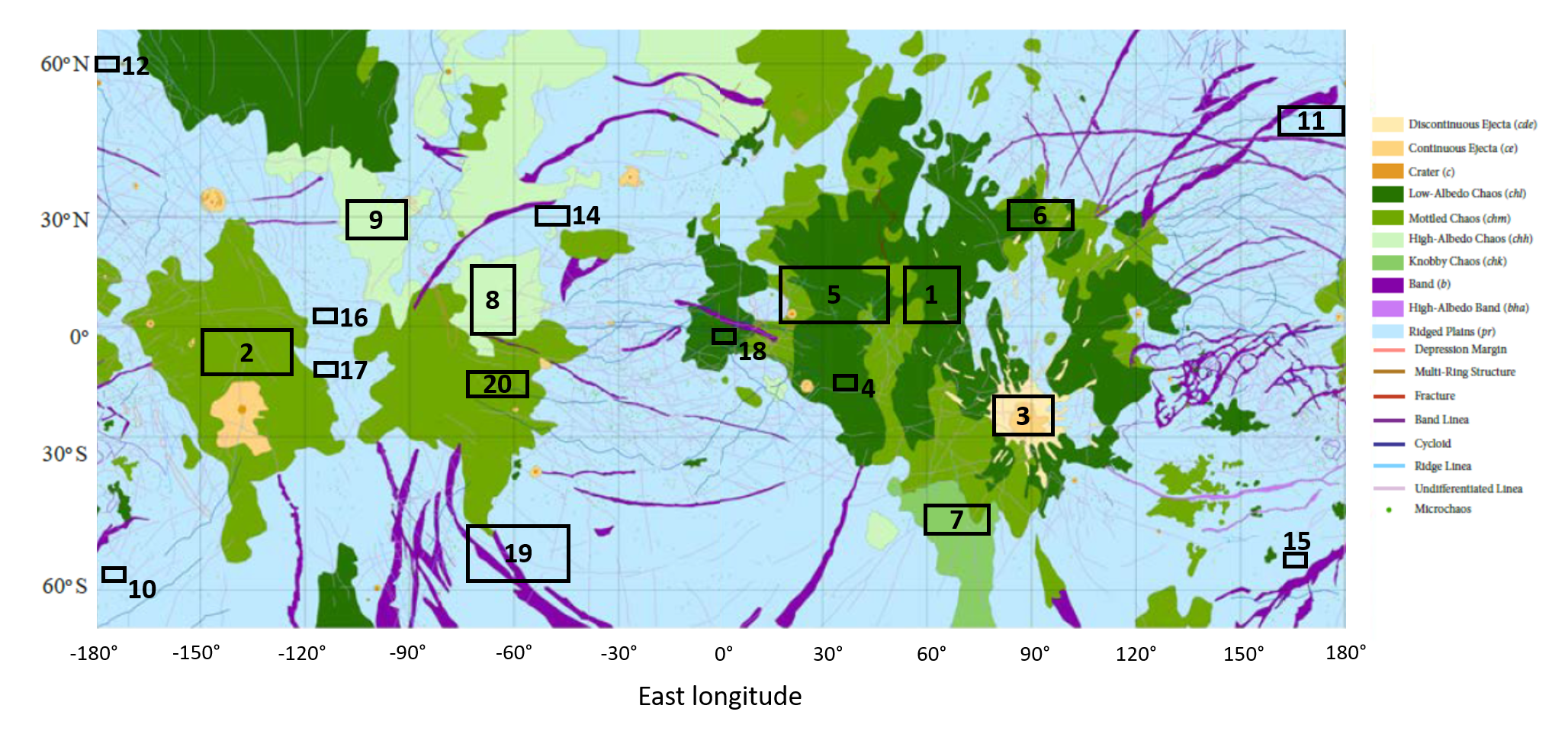}}
\caption{Geologic map of Europa \citep{Leonard_EuropaGlobalGeologic_2018} with ROIs approximately placed. The visualization of the map was adapted to our maps of parameters values.}
\label{fig:eur_geol_zones}
\end{figure}

The roughest areas are ROIs$\#$ 3, 7 and 9 (see fig. \ref{fig:eur_map_theta}). The first is the ejecta blanket of crater Pwyll, which we expect to be rough. The last two are mapped as chaos (respectively "knobby" and "high-albedo") in \citealt{Leonard_EuropaGlobalGeologic_2018} (see fig. \ref{fig:eur_geol_zones}) which is also coherent with high roughness of different scales. But we can see other areas close to the equator also mapped as chaos that show much lower values of $\overline{\theta}$.

It is challenging to have a definite interpretation of the values of $\overline{\theta}$ since it is supposed to be a combination of all the different scales from the pixel down to the individual grain \citep{Hapke_BidirectionalReflectanceSpectroscopy_I_1984}. However, it has been shown that $\overline{\theta}$ is mostly related to the smallest scale able of generating a shadow \citep{Shepard_ShadowsPlanetarySurface_I_1998} and this seems to be confirmed in more recent work \citep{Shepard_LaboratoryStudyBidirectional_I_2011,  Labarre_SurfaceRoughnessRetrieval_I_2017}.

It is also worth noting that in a number of ROIs, few data is available at phases greater than $40\degree$ where the photometric effects of the macroscopic roughness are most significant \citep{Hapke_BidirectionalReflectanceSpectroscopy_I_1984,Buratti_VoyagerPhotometryEuropa_I_1983, Buratti_ApplicationRadiativeTransfer_I_1985}.

\subsubsection*{Particle phase function}

The shape of the diffusion lobe is directly connected to the particle scattering phase function and its two parameters $b$ and $c$. The asymmetry parameter $b$ describes the isotropy ($b<0.5$) or anisotropy ($b>0.5$) of the diffusion whereas the backscattering fraction $c$ describes the main direction of the diffusion and is indicative of a forward ($c<0.5$) or backward scattering ($c>0.5$).

\begin{figure}[h!]
\centerline{\includegraphics[width=18cm]{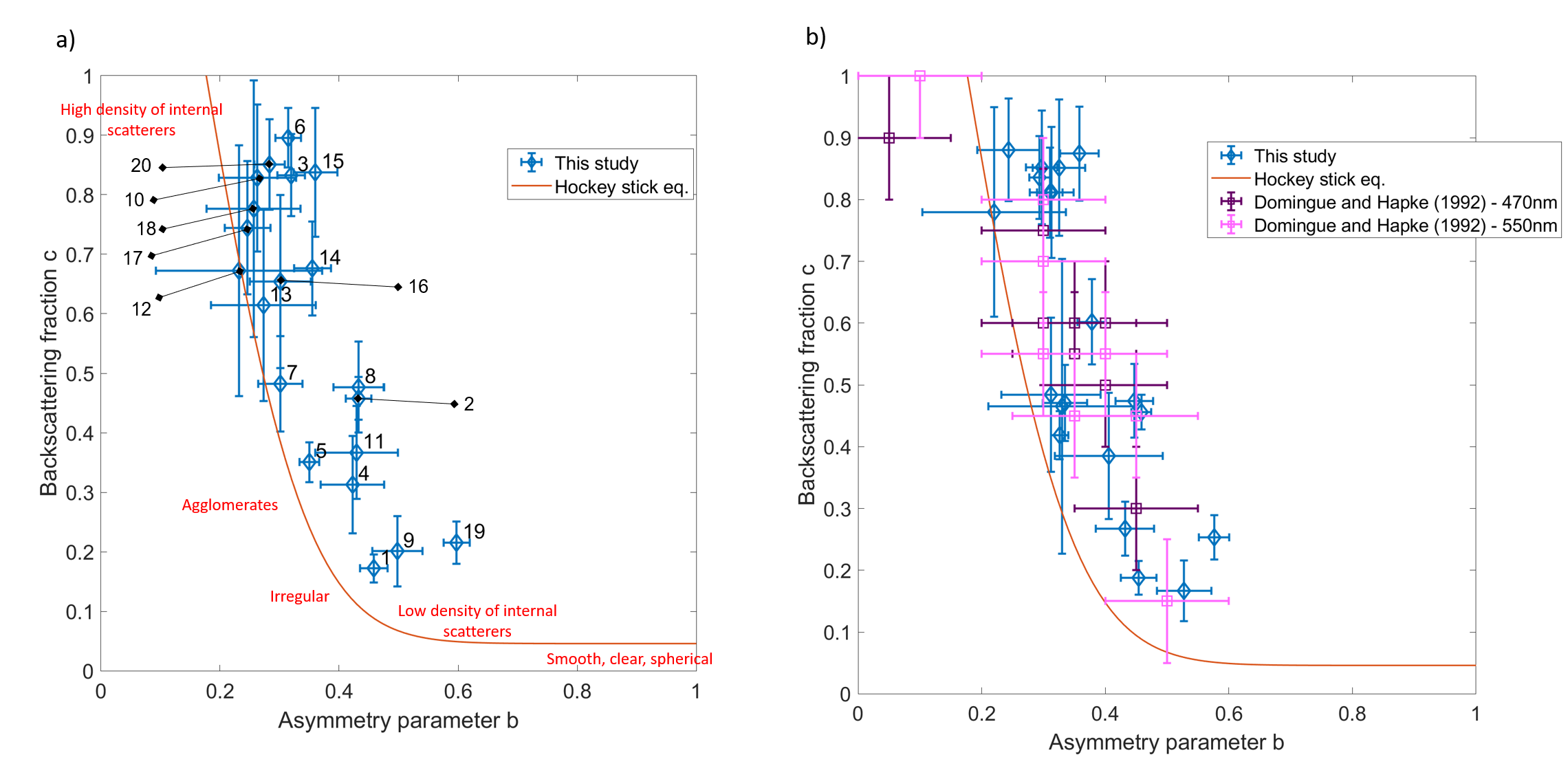}}
\caption{Particle phase function parameters $b$ and $c$. a) results of our study of Europa plotted with the Hockey stick relation of \citealt{Hapke_BidirectionalReflectanceSpectroscopy7_I_2012} and physical descriptors of the experimental study \citep{McGuire_ExperimentalStudyLight_I_1995, Souchon_ExperimentalStudyHapkes_I_2011}. b) same data but compared to disk-resolved study of  \citealt{Domingue_DiskResolvedPhotometric_I_1992}}
\label{fig:b_v_c}
\end{figure}

Fig. \ref{fig:b_v_c}a shows our results over all the different ROIs for the parameters $b$ and $c$. We note that overall, $b$ is usually better constrained than $c$. It could be a consequence of our dataset not having much data at high phase angles which can be challenging to constrain the backscattering fraction. However, we note that in some cases such as ROI$\#5$, a great number of pixels and consequently of geometries (incidence / emergence) can still contain enough information to constrain the backscattering fraction albeit a limited phase coverage.

The values of the particle phase function parameters can shed light on the microtexture of the particles at the surface of Europa in the studied ROIs as demonstrated by \citealt{McGuire_ExperimentalStudyLight_I_1995}. They carried out an experimental study of the photometry of isolated particles and the relation between their behavior and the associated values of parameters $b$ and $c$. All of them fall into an L-shaped curve in the b v. c plot and their position is directly linked to their physical properties. This behavior has been confirmed on granular material \citealt{Souchon_ExperimentalStudyHapkes_I_2011}. Even if these studies have been performed on rocks, it should also apply for ices. Experimental studies have also been conducted on water-ice analogs \citep{Jost_MicrometerSizedIce_I_2013,Jost_ExperimentalCharacterizationOpposition_I_2016,Pommerol_PhotometryBulkPhysical_PaSS_2011} also shows that forward scattering is observed in the case of small rounded particles.

Values of $b$ are mostly centered around $0.3$ describing a rather isotropic diffusion ($b<0.5$). We mostly have backscattering areas ($c>0.5$) which is consistent with disk integrated studies of Europa \citep{Buratti_ApplicationRadiativeTransfer_I_1985, Domingue_EuropasPhaseCurve_I_1991, Domingue_ReAnalysisSolar_I_1997} and the areas studied by \citealt{Domingue_DiskResolvedPhotometric_I_1992}. Moreover, if we take a closer look at fig. \ref{fig:b_v_c}a, we notice three groups of areas:
\begin{itemize}
\item a group of ROIs in the top left corner of the L-curve, very backscattering ($c \geq 0.8$) and containing most of the results
\item a group of ROIs in the middle of the L-curve, with an intermediate directional scattering ($0.3 \leq c \leq 0.8$)
\item a group of ROIs in the lowermost part of the L-curve, quite forward scattering ($c \leq 0.3$)
\end{itemize}

The first group's behavior is consistent with complex particles with a high density of internal scatterers. The second group's behavior resembles more that of agglomerates or particles of a fair number of internal scatterers. We can compare these values to laboratory measurements of micrometer-sized ice particles that showed a very distinct backscattering behavior and no forward scattering peak at all \citep{Jost_MicrometerSizedIce_I_2013}. A subsequent study by \citealt{Jost_ExperimentalCharacterizationOpposition_I_2016} explored samples with a similar behavior to a few of our backscattering areas - namely, the samples using rough solid glass spheres (called ice method $\#$3) consisting of a rough surface composed of agglomerates of fine ice particles. We should also note that this is very different from terrestrial snow which is highly forward scattering \citep{Verbiscer_ScatteringPropertiesNatural_I_1990,Domingue_ScatteringPropertiesNatural_I_1997}.  

\begin{figure}[h!]
\centerline{\includegraphics[width=16cm]{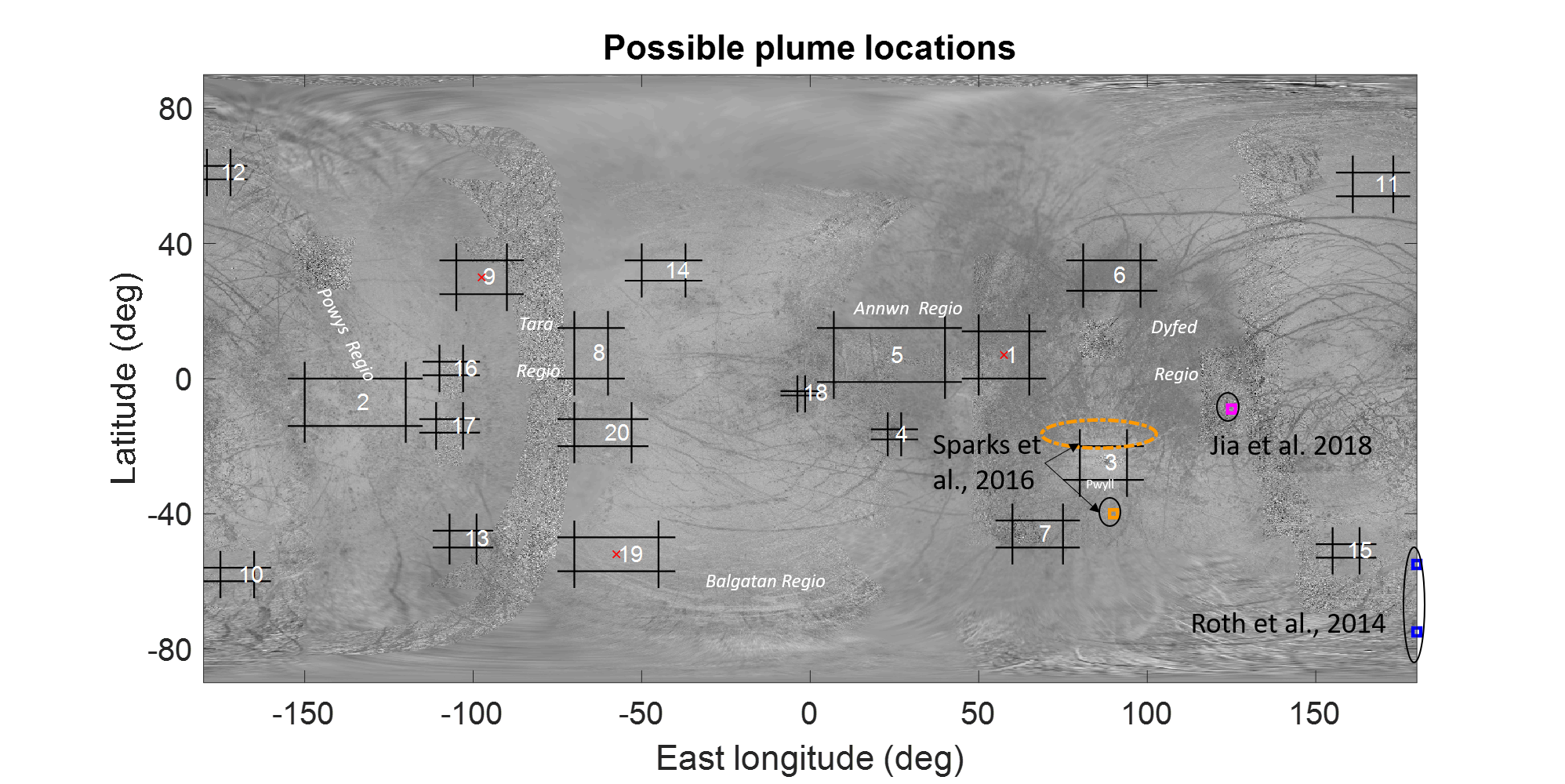}}
\caption{Background map of Europa with estimated plume locations from previous studies \citep{Roth_TransientWaterVapor_S_2014, Sparks_ProbingEvidencePlumes_AJ_2016, Jia_EvidencePlumeEuropa_NA_2018}. These locations are approximated based on modeling \citep{Roth_TransientWaterVapor_S_2014} or simulations \citep{Jia_EvidencePlumeEuropa_NA_2018}. Uncertainties are not known but should be large as observation are limb measurements. }
\label{fig:map_plumes}
\end{figure}

Finally, the third group is very forward scattering with a moderate asymmetry parameter $b$. This corresponds to particles with a low density of internal scatterers. Two areas show both forward scattering ($c<0.5$) and a narrower lobe ($b>0.5$): ROIs $\#9$ and $\#19$. Both are on the leading side of Europa where space weathering is less important \citep{WesleyPatterson_CharacterizingElectronBombardment_I_2012}. More space weathering leads to a multiplication of internal scatterers which result in a more backward scattering and isotropic diffusion lobe \citep{McGuire_ExperimentalStudyLight_I_1995, Hartman_ScatteringLightIndividual_I_1998}. They are situated between east longitudes -50$\degree$ and -100$\degree$ and respectively at latitude 25$\degree$ and -57$\degree$. We should note that this is nowhere near the previously identified candidate plume locations (see fig. \ref{fig:map_plumes}), although there is a significant uncertainty on the determination of these locations \citep{Roth_TransientWaterVapor_S_2014, Sparks_ProbingEvidencePlumes_AJ_2016, Sparks_ActiveCryovolcanismEuropa_TAJ_2017} as they are limb detections. 

These two areas (ROIs\# 9 and 19) possibly present a different kind of material than the others, such as fresh deposits from frost, cryovolcanism or maybe even the mysterious plumes that have been rescently observed in images of the Hubble Space Telescope\citep{Roth_TransientWaterVapor_S_2014, Rhoden_LinkingEuropasPlume_I_2015, Sparks_ProbingEvidencePlumes_AJ_2016} or recent cryovolcanism. They are also both very bright with a single scattering albedo of 0.99. 

ROI$\#$1 also has a similar behavior with an asymmetry parameter $b$ of 0.45. This area is on the trailing hemisphere, close to the equator. Although it is not as bright as ROIs$\#$ 9 and 19, it is still fairly bright for the trailing hemisphere with a single scattering albedo of 0.92. It could also be the presence of relatively fresh deposits with coarser grains maybe suggesting a different origin than for the forward scattering areas of the leading hemisphere.

\subsubsection*{Opposition effect}

The values of the opposition parameters are summarized in table \ref{tab:results_oppo}. We added the values of non-uniformity criterion $k$ \citep{Fernando_SurfaceReflectanceMars_JoGRP_2013} introduced at the beginning of this section. Parameters $h$ and $B_0$ are the hardest to constrain in this study even though we have a few observations below $10\degree$ phase angle. A more thorough characterization of the opposition surge would require a lot more data closer to $0\degree$ phase angle as demonstrated by studies on Europa and other bodies \citep{Domingue_EuropasPhaseCurve_I_1991, Verbiscer_OppositionSurgeEnceladus_I_2005,  Buratti_DiscoveryRemarkableOpposition_LaPSC_2019}.

\begin{figure}[h!]
\centering
	\begin{subfigure}[h!]{0.71\textwidth}
	\includegraphics[width=\textwidth, trim={0cm 1cm 0cm 0.5cm}, clip=true]{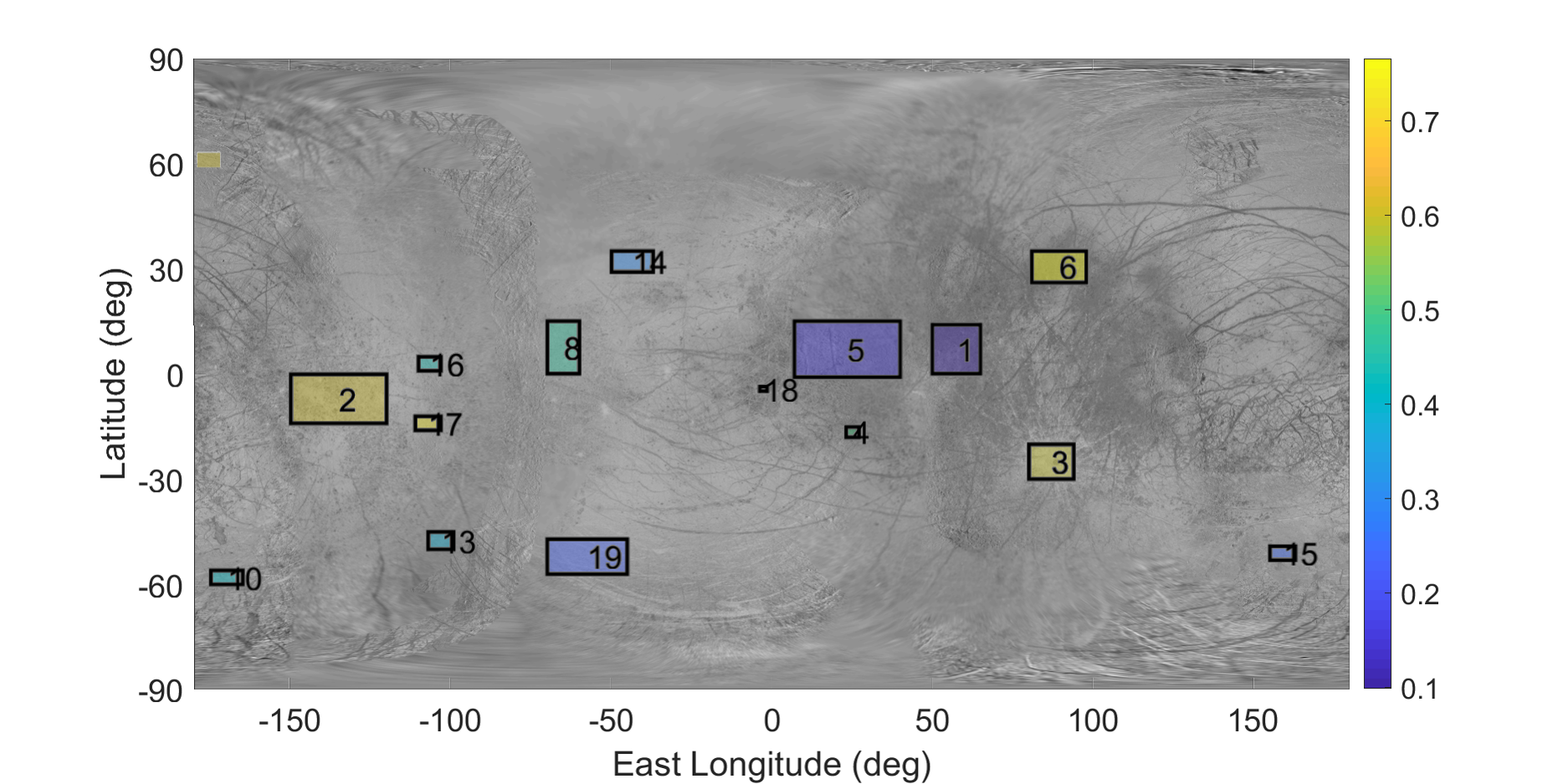}
	\subcaption{Values of the opposition surge half width $h$}
	\label{fig:eur_map_h}
	\end{subfigure}

	\begin{subfigure}[h!]{0.71\textwidth}
	\includegraphics[width=\textwidth, trim={0cm 1cm 0cm 0.5cm}, clip=true]{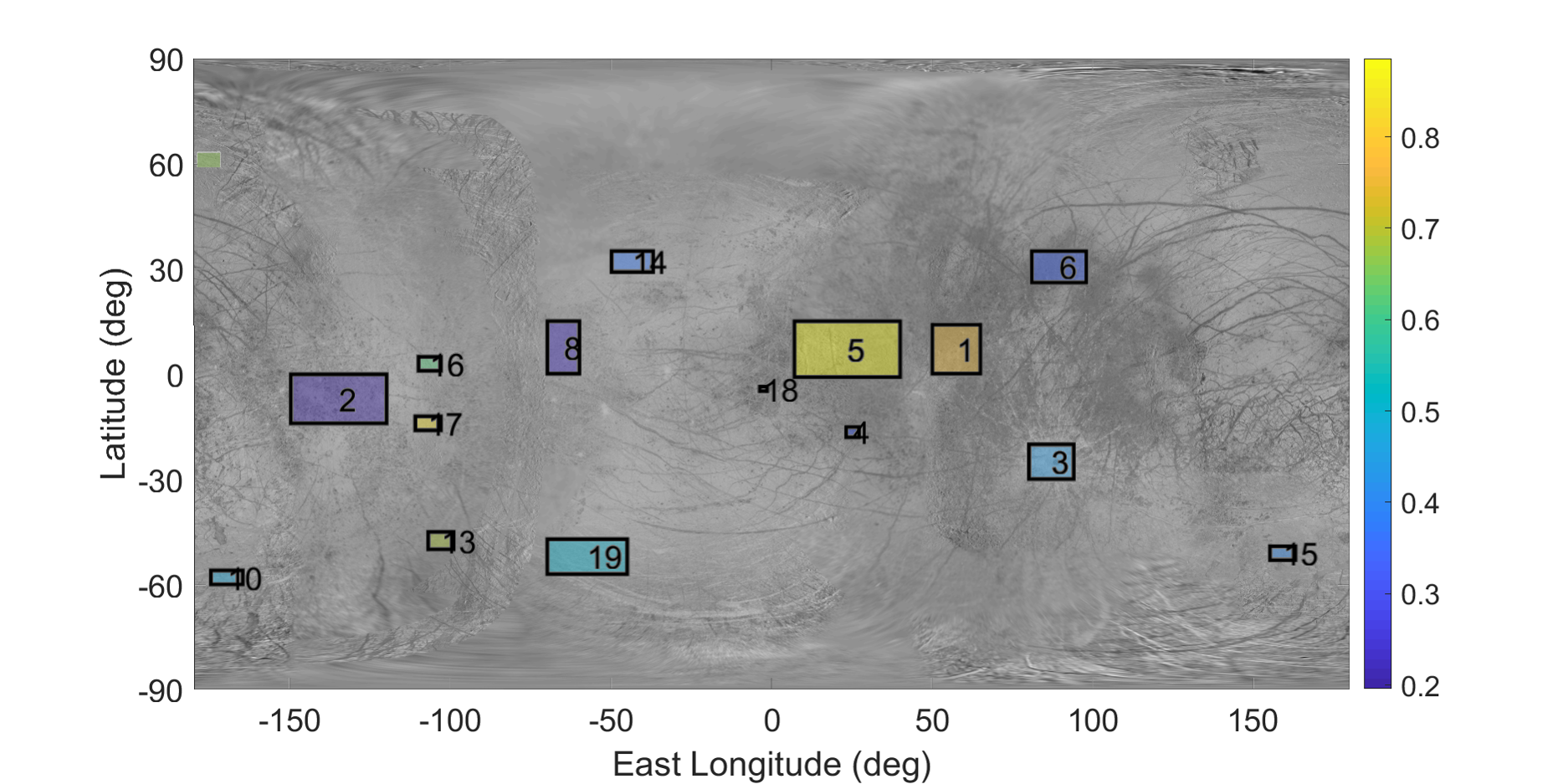}
	\subcaption{Values of the opposition surge amplitude $B_0$}
	\label{fig:eur_map_B0}
	\end{subfigure}
		
	\caption{Value of the estimated opposition effect parameters across the surface of Europa (mean of a posteriori PDF) for areas verifying k>0.5 for both $h$ and $B_0$}
	\label{fig:eur_hapke_maps_oppo}
\end{figure}

$h$ and $B_0$ are correlated so we need to look at areas where both their a posteriori PDFs are non-uniform ($k>0.5$). This the case of most of our areas (in white in table \ref{tab:results_oppo}) but we need to be careful in our interpretation as even areas where the PDFs are non-uniform can not be very well constrained with large uncertainties (std columns in table \ref{tab:results_oppo}). We also show the maps of parameters $h$ and $B_0$ for those areas (fig. \ref{fig:eur_hapke_maps_oppo}).

We can note that there seems to be a slight hemispherical heterogeneity with a trailing side showing a narrower opposition surge (low $h$) on average. The mean value of the amplitude $B_0$ on these ROIs is of $0.5$ which is consistent with \citealt{Domingue_EuropasPhaseCurve_I_1991}'s study who found the same value. However, our values of $h$ are a lot higher than any previous studies. For instance,  \citealt{Domingue_DiskResolvedPhotometric_I_1992} fixed it at 0.0016 based on their previous work. However, constraining such a narrow opposition surge (half width of 0.1$\degree$ would be impossible with our dataset. 

We should note that in its 1993 form, the Hapke model is only taking into account the Shadow Hiding Opposition Effect (SHOE). \citealt{Jost_ExperimentalCharacterizationOpposition_I_2016} studied in detail the opposition effect of water-ice analogs with the 2002 version on the Hapke model with a more precise description of the opposition effect. They concluded that the Coherent Backscattering Opposition Effect (CBOE) dominates on fresh deposit material but becomes less consequential as the particles become more irregular and porous. No direct comparison of the value of $B_0$ or $h$ is really possible to our results. Additionally, our dataset contains few data with phase angles below 10$\degree$ and is thus not very informative about the opposition effect.

\begin{table}[h!]
	\centering
	\begin{tabular}{|l|c|c|c|c|c|c|}
		\hline
    	~ 	& \multicolumn{3}{c}{$h$} & \multicolumn{3}{|c|}{$B_0$}
		\\
		ROI 	& mean & std & k 	& mean & std & k
		\\ \hline

		\csvreader[late after line=\\\hline,
filter expr={
      test{\ifnumgreater{\thecsvinputline}{1}}
  and test{\ifnumless{\thecsvinputline}{8}}
}]%
		{params_invhapke_parallel70_sigL=30_tol=30.csv}{ROI=\ROI,hmean=		\hmean,hstd=\hstd,heu=\heu,Bmean=\Bmean,Bstd=\Bstd,Beu=\Beu}%
{\ROI & \hmean &\hstd & \heu & \Bmean &\Bstd & \Beu}%

\rowcolor{Gray}
\csvreader[late after line=\\\hline,
filter expr={
      test{\ifnumgreater{\thecsvinputline}{7}}
  and test{\ifnumless{\thecsvinputline}{9}}
}]%
{params_invhapke_parallel70_sigL=30_tol=30.csv}{ROI=\ROI,hmean=\hmean,hstd=\hstd,heu=\heu,Bmean=\Bmean,Bstd=\Bstd,Beu=\Beu}%
{\ROI & \hmean &\hstd & \heu & \Bmean &\Bstd & \Beu}%

\csvreader[late after line=\\\hline,
filter expr={
      test{\ifnumgreater{\thecsvinputline}{8}}
  and test{\ifnumless{\thecsvinputline}{10}}
}]%
{params_invhapke_parallel70_sigL=30_tol=30.csv}{ROI=\ROI,hmean=\hmean,hstd=\hstd,heu=\heu,Bmean=\Bmean,Bstd=\Bstd,Beu=\Beu}%
{\ROI & \hmean &\hstd & \heu & \Bmean &\Bstd & \Beu}%

\rowcolor{Gray}
\csvreader[late after line=\\\hline,
filter expr={
      test{\ifnumgreater{\thecsvinputline}{9}}
  and test{\ifnumless{\thecsvinputline}{11}}
}]%
{params_invhapke_parallel70_sigL=30_tol=30.csv}{ROI=\ROI,hmean=\hmean,hstd=\hstd,heu=\heu,Bmean=\Bmean,Bstd=\Bstd,Beu=\Beu}%
{\ROI & \hmean &\hstd & \heu & \Bmean &\Bstd & \Beu}%

\rowcolor{Gray}
\csvreader[late after line=\\\hline,
filter expr={
      test{\ifnumgreater{\thecsvinputline}{10}}
  and test{\ifnumless{\thecsvinputline}{12}}
}]%
{params_invhapke_parallel70_sigL=30_tol=30.csv}{ROI=\ROI,hmean=\hmean,hstd=\hstd,heu=\heu,Bmean=\Bmean,Bstd=\Bstd,Beu=\Beu}%
{\ROI & \hmean &\hstd & \heu & \Bmean &\Bstd & \Beu}%

\rowcolor{Gray}
\csvreader[late after line=\\\hline,
filter expr={
      test{\ifnumgreater{\thecsvinputline}{11}}
  and test{\ifnumless{\thecsvinputline}{13}}
}]%
{params_invhapke_parallel70_sigL=30_tol=30.csv}{ROI=\ROI,hmean=\hmean,hstd=\hstd,heu=\heu,Bmean=\Bmean,Bstd=\Bstd,Beu=\Beu}%
{\ROI & \hmean &\hstd & \heu & \Bmean &\Bstd & \Beu}%

\csvreader[late after line=\\\hline,
filter expr={
      test{\ifnumgreater{\thecsvinputline}{12}}
  and test{\ifnumless{\thecsvinputline}{21}}
}]%
{params_invhapke_parallel70_sigL=30_tol=30.csv}{ROI=\ROI,hmean=\hmean,hstd=\hstd,heu=\heu,Bmean=\Bmean,Bstd=\Bstd,Beu=\Beu}%
{\ROI & \hmean &\hstd & \heu & \Bmean &\Bstd & \Beu}%

\rowcolor{Gray}
\csvreader[late after line=\\\hline,
filter expr={
      test{\ifnumgreater{\thecsvinputline}{20}}
  and test{\ifnumless{\thecsvinputline}{22}}
}]%
{params_invhapke_parallel70_sigL=30_tol=30.csv}{ROI=\ROI,hmean=\hmean,hstd=\hstd,heu=\heu,Bmean=\Bmean,Bstd=\Bstd,Beu=\Beu}%
{\ROI & \hmean &\hstd & \heu & \Bmean &\Bstd & \Beu}%

\end{tabular}

	\caption{Value of the estimated opposition effect parameters over all the ROIs on Europa (in gray, ROIs for which either one of the opposition parameters verifies $k < 0.5$)}
	\label{tab:results_oppo}
\end{table}

\section{Discussion and conclusion}

We have performed a regional photometric study of Europa using New Horizons' LORRI and Voyager's ISS instruments on 20 Region of Interest (ROIs). We corrected the geometrical metadata and the absolute level discrepancies between the 57 images used. The estimation of the Hapke parameters in the Bayesian framework gives satisfying results.

We have found that despite the relatively small phase angle range, some ROIs have very accurate results on the $c$ parameter. This behavior is unexpected because usually phase angles higher than 90$\degree$ are required for Emission Phase Function (EPF) as demonstrated by \citealt{Schmidt_RealisticUncertaintiesHapke_I_2015} and \citealt{Schmidt_EfficiencyBRDFSampling_I_2019}. We confirmed on synthetic data that the Bayesian method is able to estimate forward and backward scattering behaviors in such limited geometries. We attribute this behavior to the fact that each phase is observed several times (several pixels of the very same images), each at various incidence/emergence angles. This information is not present in standard EPF but is definitely present in our dataset.

Most of the 20 regions are coherent with the typical bright backscattering behavior of Europa. We attribute this typical behavior due to space weathering that plays a major part on Europa especially in darkening the trailing hemisphere \citep{Paranicas_ElectronBombardmentEuropa_GRL_2001, Paranicas_EuropasRadiationEnvironment_book_2009}. Impact gardening can contribute to mixing the upper layers of the surface over time \citep{Dalton_ExogenicControlsSulfuric_PaSS_2013} and thus change their physical properties. This can contribute to form more complex particles or even aggregates that can account for more backscattering particles.

We have found that the roughness parameter is more heterogeneous than previously measured - ($\overline{\theta}$ ranges from $6\degree$ to $27\degree$). A recent study showed that penitentes could exist in an equatorial belt at latitudes below $23\degree$ \citep{Hobley_FormationMetreScale_NG_2018}. They are meter-scale blade-like structures of ice. If they are present, we would expect to see high values of roughness close to the equator. This is not what we see in our results. However, it is challenging to have a definite interpretation of the values of $\overline{\theta}$ since it is defined to be a combination of all the different scales from the pixel down to the individual grain \citep{Hapke_BidirectionalReflectanceSpectroscopy_I_1984}. However, it has been shown that $\overline{\theta}$ is mostly related to the smallest scale able of generating a shadow \citep{Shepard_ShadowsPlanetarySurface_I_1998} and this seems to be confirmed in more recent work \citep{Shepard_LaboratoryStudyBidirectional_I_2011,  Labarre_SurfaceRoughnessRetrieval_I_2017}. This means that the presence of penitentes is not necessarily measurable with values of $\overline{\theta}$. It is also worth noting that in a number of ROIs, few data is available at phases greater than $40\degree$ where the photometric effects of the macroscopic roughness are most significant \citep{Hapke_BidirectionalReflectanceSpectroscopy_I_1984,Buratti_VoyagerPhotometryEuropa_I_1983, Buratti_ApplicationRadiativeTransfer_I_1985}.

We have identified two very bright areas on the leading hemisphere ($\omega$=0.99) showing a narrow forward scattering behavior ($b\gtrapprox 0.5$ $\&$ $c<0.5$) between -100$\degree$ and -50$\degree$ longitude. This could be indicative of fresh frost deposits and make these regions of particular interest as a study site for future missions. We can speculate that this peculiar forward scattering may be due to plume activity \citep{Roth_TransientWaterVapor_S_2014,  Rhoden_LinkingEuropasPlume_I_2015, Sparks_ProbingEvidencePlumes_AJ_2016} which could result in fresh material deposited on the surface. Fresh water ice deposits would tend to be brighter than older terrains, as would the presence of small grains. This is both consistent with our results and the presence of plume deposits. However, we have to note that the most probable region identified on the Hubble Space Telescope observations as the source of plumes is near the Pywll crater. This is ROI$\#$3 that has a decidedly backscattering behavior. Another possibility is the presence of recent cryovolcanism that could expose fresh water-ice from below the icy crust and be responsible for a number of lobate features such as lenticulae and chaos \citep{Miyamoto_PutativeIceFlows_I_2005, Quick_ConstraintsDetectionCryovolcanic_PaSS_2013}. These effects would lead to younger deposits that would results in more forward scattering particles. Because ROI$\#$9 has also a high roughness (23$\degree$) and is mapped as a "high-albedo chaos" \citep{Leonard_EuropaGlobalGeologic_2018}, another possible explanation for this behavior would be the presence of disrupted young terrains with exposed endogenic particles sampled from the subsurface ocean \citep{Ligier_VLTSINFONIObservationsEuropa_TAJ_2016, Trumbo_SodiumChlorideSurface_SA_2019}. We have also identified forward scattering in ROI$\#$1 in the trailing hemisphere, although it is a lot less bright ($\omega=0.92$). Fresh deposits could also be the cause of that behavior but maybe of a different nature than in ROIs$\#$ 9 and 19. 

ROI $\#$3 is the ejecta blancket of crater Pwyll at the center of the thermal anomaly \citep{Spencer_TemperaturesEuropaGalileo_S_1999}. It seems that this region has an relatively high albedo ($\omega=0.92$) and high roughness ($\overline{\theta}=27\degree$), as expected for a recent ejecta blancket. Nevertheless, the particle phase function is not out of the general trend of an isotropic backward scattering. Thus, this region does not appear to be a very good candidate for fresh deposits resulting from plume activity.

Several points must be raised about the methodology of this study. 

First, the Hapke model we chose to work with has been discussed at length. In particular, \citealt{Shepard_TestHapkePhotometric_JoGR_2007} have shown the limitations of the physical interpretation of $\overline{\theta}$ as it is predominantly dependent on the smallest scale. Additionally, Hapke has altered his model continuously by taking into account the Coherent Backscatter Opposition Effect (CBOE) for example \citep{Hapke_BidirectionalReflectanceSpectroscopy5_I_2002, Hapke_BidirectionalReflectanceSpectroscopy6_I_2008, Hapke_BidirectionalReflectanceSpectroscopy7_I_2012}. However, as it has been widely used in its 1993 form for experimental studies and for planetary bodies, including the icy moons, it has the merit of possible comparisons to properties of analogs and other icy bodies. Future work should integrate more precise radiative transfer model, such as \citealt{Doute_MultilayerBidirectionalReflectance_JoGRP_1998} or \citealt{Andrieu_RadiativeTransferModel_AO_2015}.

Second, we have assumed that the arbitrarily defined ROIs were photometrically coherent i.e. that the entire ROI area has the same photometric behavior. A great diversity of terrains on Europa is possible even beyond our scale of observation. In the future more resolved data with a large geometry coverage is required to make complete and precise photometric maps.

Nevertheless, the whole pipeline we have developed is adapted to any additional dataset and photometric model. We plan to look for other models to accommodate the most extreme geometries explicitly removed from this study ($e,i>70\degree$).

\section*{Acknowledgements}
We would like to thank the Voyager's ISS and New Horizons' LORRI teams for making their data available on NASA's PDS. We would also like to thank Andrew Cheng, LORRI's PI for answering our questions about the pivot wavelength and Mark Showalter for his insight on the Voyager dataset. Finally, we would like to thank an anonymous reviewer for their helpful comments that improved our discussion. 
This work is supported by Airbus Defence $\&$ Space, Toulouse (France) as well as the "IDI 2016" project funded by the IDEX Paris-Saclay, ANR-11-IDEX-0003-02. We also acknowledge support from the ``Institut National des Sciences de l'Univers'' (INSU), the Centre National de la Recherche Scientifique (CNRS) and Centre National d'Etudes Spatiales (CNES) through the Programme National de Planétologie.

\bibliographystyle{elsarticle-harv}
\bibliography{HapkeEuropa_bibtex}

\end{document}